\documentclass[aps,prd,twocolumn,showpacs,floatfix,
               superscriptaddress,amsmath]{revtex4}
\usepackage{bm}
\usepackage{latexsym}
\usepackage{graphicx}
\usepackage{youngtab}
\usepackage{comment}
\def\simge{
    \mathrel{\rlap{\raise 0.511ex
        \hbox{$>$}}{\lower 0.511ex \hbox{$\sim$}}}}
\def\simle{
    \mathrel{\rlap{\raise 0.511ex
        \hbox{$<$}}{\lower 0.511ex \hbox{$\sim$}}}}
\def\beqn{\begin{equation}}
\def\eeqn{\end{equation}}
\def\barr{\begin{eqnarray}}
\def\earr{\end{eqnarray}}
\def\bc{\begin{center}}
\def\ec{\end{center}}
\renewcommand{\o}[1]{\overline{#1}}
\renewcommand{\l}{\left}
\renewcommand{\r}{\right}
\newcommand{\nn}{\nonumber}
\newcommand{\eps}{\epsilon}
\newcommand{\ot}{\otimes}
\newcommand{\op}{\oplus}
\newcommand{\om}{\ominus}
\begin{document}


\title{Clebsch-Gordan Construction of Lattice Interpolating Fields
for Excited Baryons}

\author{Subhasish~Basak}
 \affiliation{Department of Physics, University of Maryland, College Park,
              MD 20742, USA}

\author{Robert~Edwards}
 \affiliation{Thomas Jefferson National Accelerator
              Facility, Newport News, VA 23606, USA}


\author{George~T.~Fleming}
 \affiliation{Department of Physics, Yale University,
              New Haven, CT 06511, USA}

\author{Urs~M.~Heller}
 \affiliation{American Physical Society, One Research Road,
    Ridge, NY 11961-9000, USA}

\author{Colin~Morningstar}
 \affiliation{Department of Physics, Carnegie Mellon
    University, Pittsburgh, PA 15213, USA}

\author{David~Richards}
 \affiliation{Thomas Jefferson National Accelerator
              Facility, Newport News, VA 23606, USA}

\author{Ikuro~Sato}
 \affiliation{Department of Physics, University of Maryland, College Park,
              MD 20742, USA}
\author{Stephen~J.~Wallace}
 \affiliation{Department of Physics, University of Maryland, College Park,
              MD 20742, USA}

\collaboration{Lattice Hadron Physics Collaboration (LHPC)}

\pacs{11.15.Ha, 
      12.39.Mk  
      12.38.Gc  
}

\date{\today}

\begin{abstract}
Large sets of baryon interpolating field operators are developed
for use in lattice QCD studies of baryons with zero momentum.
Operators are classified according to the double-valued
irreducible representations of the octahedral group. At first,
three-quark smeared, local operators are constructed for each
isospin and strangeness and they are classified according to
their symmetry with respect to exchange of Dirac indices. Nonlocal
baryon operators are formulated in a second step as direct
products of the spinor structures of smeared, local operators
together with gauge-covariant lattice displacements of one or
more of the smeared quark fields. Linear combinations of direct
products of spinorial and spatial irreducible representations are
then formed with appropriate Clebsch-Gordan coefficients of the
octahedral group. The construction attempts to maintain maximal
overlap with the continuum $SU(2)$ group in order to provide a
physically interpretable basis.  Nonlocal operators provide
direct couplings to states that have nonzero orbital angular
momentum.
\end{abstract}

\maketitle

\section{Introduction}
\label{sec:intro}

The theoretical determination of the spectrum of baryon resonances
from QCD continues to be an important goal.  Lattice QCD
calculations have succeeded in part to meet this goal by providing
results for the lowest-mass baryon of each isospin in the quenched
approximation using overly large masses for the $u$ and $d$
quarks.~\cite{Aoki03,Weingarten93}

Most lattice simulations to date have used restricted sets of
operators appropriate for $J^P = 1/2^\pm, 3/2^\pm$ states.  Masses
of low-lying, positive-parity baryons are reproduced with
approximately 10\% deviations from experimental values using the
quenched approximation~\cite{Aoki03}.   Much less is known about
higher-spin excited states. The first preliminary lattice
calculation of $5/2^\pm$ $N^*$ masses was reported by the Lattice
Hadron Physics Collaboration~\cite{Basak04} using one of the
operators that we develop in this paper~\cite{Sato04}.  Results
for excited baryons were also reported based on the  use of
different radial smearings of the quarks in
Refs.~\cite{BGR04,BGR05}. Recently, studies of negative-parity
baryons have been reported by several
groups~\cite{Kyoto03,Gockeler02,Sasaki05,Sasaki02,
Adelaide03,Leinweber03,Regensburg03}. Nemoto {\it et al.} and
Melnitchouk {\it et al.} considered the $\Lambda(1405)$ baryon,
which is the lightest negative-parity baryon despite its nonzero
strangeness. They used a three-quark interpolating field operator
motivated by the spin-flavor $SU(6)$ quark model and concluded
that $\Lambda(1405)$ was not evident in their lattice
calculations.

To improve upon our understanding
of the resonance spectrum, correlation matrices will be needed,
necessitating the construction of sufficiently large sets of
baryon and multi-hadron operators.
The correlation-matrix method~\cite{michael85,lw90} has been used to
determine the spectrum of glueball masses by Morningstar and
Peardon~\cite{Morningstar99}. A large number of interpolating
field operators was used to form matrices of lattice
correlation functions. The spectrum of effective masses was
obtained by diagonalizing the matrices of correlation functions
to isolate mass eigenstates for each symmetry channel. In effect, one
allows the dynamics to determine the optimal linear combination of
operators that couple to each mass state. A similar program for
baryons is being undertaken by the Lattice Hadron Physics
Collaboration.  The first step is to determine a large number of
suitable baryon interpolating field operators that correspond to
states of zero momentum, definite parity and values of angular
momentum $J = {1\over 2}, {3 \over 2}, {5 \over 2}, \cdots$.

Due to the complexity of the operator construction and the
importance of providing checks on the final results, we have been
pursuing two different approaches. An alternative method, based
on a computational implementation of the group projection
operation, is presented elsewhere~\cite{Morningstar04,Basak05}.

On a cubic lattice, the continuum rotational symmetry is broken
to the finite octahedral group, $O$. States of definite angular
momentum correspond to states that occur in certain patterns
distributed over the irreducible representations (IRs) of $O$.
Although mass calculations are insensitive to the spin projection
$J_z$, other applications do require baryons with a definite spin
projection. We develop operators that are IRs of $O$ using a
basis that corresponds as closely as possible to the continuum
$J,J_z$ states in order to have operators for applications that
require spin projection.

It is important to use smearing of the quark fields and to have
nonlocal baryon operators as well as the usual local operators.
Smeared and nonlocal operators provide a variety of radial and
angular distributions of quarks within a baryon so as to couple
efficiently to excited states. Nonlocal operators are needed in
order to realize spins $J > { 3\over 2}$ and simply to enlarge
the sets of operators.

In this paper, we first review some basic facts about the
octahedral group for integer and half-integer spins in
Section~\ref{sec:grouptheory}~\cite{Johnson,Elliott,Altmann,Mandula83,
Mandula82,Butler} and review a useful notation for Dirac indices
based on $\rho$-spin. Two basic types of three-quark operators are
considered: quasi-local and nonlocal. Each quark field in a
quasi-local baryon operator is smeared about a common point ${\bf
x}$ using the same cubically symmetric form of smearing.
Quasi-local operators include local operators as a special case,
i.e., when the smearing is omitted.  We develop IRs for
quasi-local operators in Section~\ref{sec:local} for each baryon:
$N$, $\Delta$, $\Lambda$, $\Sigma$, $\Xi$ and $\Omega$. This
amounts to determining all allowed combinations of Dirac indices
for each flavor symmetry and classifying them into IRs of the
octahedral group.

The quasi-local operators provide templates that are used for the
construction of nonlocal operators in Section~\ref{sec:nonlocal}.
Nonlocal operators are formed by applying extra lattice
displacements to one (or more) of the smeared quark fields, thus
providing a smearing distribution that differs from that used for
the other quarks. In the simplest case, the combination of extra
displacements used is cubically symmetric and only changes the
radial distribution of the smearing. In other cases the
combinations of the extra displacements transform as IRs of the
octahedral group, and are chosen to correspond as closely as
possible to spherical harmonics. The IRs of lattice displacements
and IRs of Dirac indices are combined to form overall IRs for the
nonlocal operators using an appropriate set of Clebsch-Gordan
coefficients of the octahedral group. Some concluding remarks are
presented in Section~\ref{sec:conclusions}.

\section{Octahedral Group and Lattice Operators}
\label{sec:grouptheory}

In lattice QCD, hadron field operators are composed of quark and
gluon fields on a spatially-isotropic cubic lattice.  The lattice
is symmetric with respect to a restricted set of rotations about
spatial axes that form the octahedral group, $O$, which is a
subgroup of the continuum rotational group $SU(2)$. The octahedral
group consists of 24 group elements, each corresponding to a
discrete rotation that leaves invariant a cube, or an octahedron
embedded within the cube. When the objects that are rotated
involve half-integer values of the angular momentum, the number
of group elements doubles to extend the range of rotational
angles from $2\pi$ to $4\pi$, forming the double-valued
representations of the octahedral group, referred to as $O^D$.

Spatial inversion commutes with all rotations and together with
the identity forms a two-element point group. Taking inversion
together with the finite rotational group simply doubles the
number of group elements, giving the group $O^D_h$ for
half-integer spins.


Given a lattice interpolating operator for a baryon, one may
generate other operators by applying the elements of $O^D_h$ to
the given operator. This produces a set of operators that
transform amongst themselves with respect to $O^D_h$, and thus
these operators ${\cal O}_i$ form the basis of a representation
of the group. When a group element $G_a$ is applied to operator
${\cal O}_i$ in the set, the result is a linear combination of
other operators in the set, $\sum_{j}{\cal O}_j T_{ji}(G_a)$,
where $T_{ji}(G_a)$ is a matrix representation of the octahedral
group. Such matrix representations are in general reducible. In
order to identify operators that correspond to baryons with
specific lattice symmetries, it is necessary to block-diagonalize
$T_{ji}$, each block corresponding to an irreducible
representation of the octahedral group.  This task is facilitated
by a judicious choice of IR basis vectors for the octahedral
group, such as the  ``cubic harmonics'' or ``lattice harmonics''
of Refs.~\cite{Cracknell,Dirl}.

\subsection{Integer angular momentum : $O$ }
\label{subsec:O}

The octahedral group $O$ has five IRs, namely $A_1, A_2, E, T_1$
and $T_2$ with dimension 1, 1, 2, 3 and 3, respectively, where we
follow the conventions of Ref.~\cite{Johnson}.
 The patterns of IRs of $O$ that
correspond to IRs of the continuum rotational group $SU(2)$
with spin $J$ are shown in Table~\ref{table:subduction_O}.
\newcommand{\spclA}{$\oplus~n(A_1 \oplus A_2 \oplus 2E\oplus 3T_1\oplus 3T_2)$}
\begin{table}[h]
\caption{The subduction of $SU(2)$ to IR $\Lambda$ of $O$ for
integer $J$. }
\begin{tabular}{|cll|}
\hline
~~$J$~~ &  &  ~~~~~~$\Lambda$ \\
\hline
\hline
$ 0 $& & $A_1 $  \\
$ 1 $& & $T_1 $ \\
$ 2 $& & $E \op T_2 $ \\
$ 3 $& & $A_2 \op T_1 \op T_2 $ \\
$ 4 $& & $A_1 \op E \op T_1\op T_2$ \\
\hline
\end{tabular}
\label{table:subduction_O}
\end{table}
 A $J = $0 state must show up in the $A_1$
IR, but in no other IR of $O$. A $J=$ 1 state must show up
in the $T_1$ IR but in no other IR. A $J=$ 2 state must show
up in the $E$ and $T_2$ IRs.

Lattice displacements form representations of $O$ corresponding
to integer angular momenta.  We choose the standard ``lattice
harmonics" that are shown in Table~\ref{table:lattice_harmonics}
as the appropriate basis vectors because they have a
straightforward connection to IRs of the rotation group $SU(2)$ in
the continuum limit.
\begin{table}[h]
\caption{Basis of irreducible representations of $O$ in terms of
spherical harmonics, $Y_{l,m}$ for the lowest values of $l$.
$d_\Lambda$ is the dimension of the IR. The lattice harmonics are
understood to be evaluated on a cubic lattice. }
\begin{ruledtabular}
\begin{tabular}{lcccc}
$\Lambda$ & $d_\Lambda$ & row 1 & row 2 &  row 3~~ \\
\hline
$A_1$ & 1 & $Y_{0,0}$              &           $-$          & $-$    \\
$A_2$ & 1 & ~~${1\over \sqrt{2}}(Y_{3,2}-Y_{3,-2})$~~ & $-$ & $-$    \\
$E$   & 2 & $Y_{2,0}$& ~~${1\over\sqrt{2}}(Y_{2,2}+Y_{2,-2})$~~ & $-$    \\
$T_1$ & 3 & $Y_{1,1}$              & $Y_{1,0}$              & $Y_{1,-1}$ \\
$T_2$ & 3 & $Y_{2,1}$ & ${1\over\sqrt{2}}(Y_{2,2}-Y_{2,-2})$& $Y_{2,-1}$ \\
\end{tabular}
\end{ruledtabular}
\label{table:lattice_harmonics}
\end{table}
For example, the $Y_{1,m}$ spherical harmonics for $m=1, 0, -1$
provide a basis for the three-dimensional $T_1$ IR. The same
basis convention for $T_1$ appears in Ref.~\cite{Wingate95}, and
the same basis convention for $E$ appears in Ref.~\cite{Lacock96}.

 Any quantities that transform in the same fashion as the basis vectors provide a
suitable IR for the octahedral group.  We will show in
Section~\ref{sec:nonlocal} how to use combinations of lattice
displacements of quark fields in order to realize the same
transformations as the ``lattice harmonics''.
\subsection{Half-integer angular momenta: $O^D$}
\label{subsec:OD}

 The eight IRs of the double-valued representations of the
octahedral group, $O^D$, include $A_1, A_2, E, T_1, T_2$ for
integer spins and $ G_1, G_2$, and $H$ for half-integer spins.
The additional IRs $G_1, G_2$ and $H$ have dimensions 2, 2, and 4,
respectively; these are the appropriate IRs for baryon operators
on a cubic lattice.
Table~\ref{table:subduction_OD} shows the patterns within $O^D$
that correspond to some half-integer values of $J$.
\newcommand{\spclB}{$\op~2n(G_1 \op 2H \op G_2)$}
\begin{table}[h]
\caption{The subduction of $SU(2)$ to IR $\Lambda$ of $O^D$ for
half-integer $J$.}
\label{table:subduction_OD}
\begin{center}
\begin{tabular}{|cll|}
\hline
~~~$J$~~~ &  & ~~~$\Lambda$ \\
\hline
\hline
$1/2$& & $G_1 $              \\
$3/2$& & $H $                \\
$5/2$& & $H \op G_2 $        \\
$7/2$& & $G_1\op H \op G_2$  \\
 \hline
\end{tabular}
\end{center}
\end{table}
 For example, a $J =$1/2 baryon state should show up in IR $G_1$.
 A spin 3/2 baryon should show up in IR $H$. A spin 5/2 state should show up in IRs
$H$ and $G_2$ but not in $G_1$. A $J = $7/2 state should show up
once in IRs $G_1$, $H$ and $G_2$.

A suitable set of IR basis vectors for half-integer angular
momenta is given by the eigenstates $\l| J,m \r>$ of $J^2$ and
$J_z$ that are listed in Table~\ref{tab:spin_harmonics}.
\begin{table}[h]
\caption{Correspondence of our choice of rows in the $G_1$, $G_2$
and $H$ IRs to the eigenstates $\l| J,m \r>$ of $J^2$ and $J_z$. }
\begin{ruledtabular}
\begin{tabular}{cccc}
$\Lambda$ & $d_\Lambda$ & row 1 & row 2 \\
\hline
$G_1$ & 2 & $\l| {1\over 2},{1\over2} \r>$ & $\l| {1\over 2},-{1\over2} \r>$
\\
$G_2$ & 2 & \hspace{-2mm} $\sqrt{\frac{1}{6}} \l| {5\over 2}, {5\over2} \r> -
        \sqrt{\frac{5}{6}} \l| {5\over 2},-{3\over2} \r>$ &
    $-\sqrt{\frac{5}{6}} \l| {5\over 2},{3\over2} \r> +
    \sqrt{\frac{1}{6}} \l| {5\over 2},-{5\over2} \r>$
\end{tabular}
\vspace{3mm}
\begin{tabular}{cccccc}
$\Lambda$ & $d_\Lambda$ & row 1 & row 2 & row 3 & row 4 \\
\hline
$H$   & 4 & $\l| {3\over 2},{3\over2} \r>$ & $\l| {3\over 2},{1\over2} \r>$
      & ~$\l| {3\over 2},-{1\over2} \r>$ & $\l| {3\over 2},-{3\over2} \r>$ \\
\end{tabular}
\end{ruledtabular}
\label{tab:spin_harmonics}
\end{table}
Explicit forms of the $G_1$ and $H$ basis states for products of
three Dirac spinors are given in
Appendix~\ref{app:DiracSymmetry}. Note that the $G_2$ basis
cannot be built using three Dirac spinors unless orbital angular
momentum is added.
\subsection{Smearing and smearing parity }
\label{subsec:Smearing} The first step in the construction of
field operators suitable for baryons is to specify primitive
three-quark operators.  Consider a generic operator formed from
three-quark fields as follows,
\begin{equation}
\eps_{abc} \: \o{q}_{\mu_1}^{a f_1}({\bf x},t)\o{q}_{\mu_2}^{b
f_2}({\bf x},t) \o{q}_{\mu_3}^{c f_3}({\bf x},t),
\label{eq:qqq-local}
\end{equation}
where $a$, $b$ and $c$ are color indices, $f_1$, $f_2$ and $f_3$
are flavor indices and $\mu_1$, $\mu_2$ and $ \mu_3$ are Dirac
indices with values $1$ to $4$.
The operator is antisymmetrized in color by the $\eps_{abc}$
factor when the (implicit) sums over $a$, $b$ and $c$ are
performed.

The use of gauge-covariant quark-field smearing,
 such as Gaussian smearing~\cite{Alford96},
Jacobi smearing~\cite{ukqcd93} or so-called Wuppertal
smearing~\cite{Guesken90}, is important for enhancing the
coupling to the low-lying states. Gauge-link
smearing~\cite{Albanese87,Hasenfratz01,Peardon04} further reduces
the coupling to the short-wavelength modes of the theory.
 Schematically, the smearing replaces each unsmeared field by
a sum of fields with a distribution function as follows,
\begin{equation}
q_\mu ({\bf x},t) = \sum_{y} \hat{G}({\bf x},{\bf x}+ {\bf y})
\tilde{q}_\mu({\bf x}+ {\bf y},t), \label{eq:GaussianSmearing}
\end{equation}
where $\tilde{q}_\mu({\bf x},t)$ denotes an unsmeared field at
point ${\bf x}$ and the smearing distribution function $\hat{G}$
is gauge covariant. When the smearing distribution is cubically
symmetric about point ${\bf x}$ and is the same for each quark
field, the baryon operator of Eq.~(\ref{eq:qqq-local}) is
referred to as quasi-local. Quasi-local operators have the same
transformations under the octahedral group as unsmeared operators.

    Nonlocal operators differ because the smearing
distribution of one or more quark fields is altered by extra
lattice displacements.  An example is the covariant derivative
formed by a linear combination of two displacements of a smeared
quark field,
\begin{equation}
D_i q_{\mu}({\bf x},t) = U_i({\bf x}) q_{\mu}({\bf x} + \hat{i},
t) - U^{\dag}_i({\bf x} - \hat{i}) q_{\mu}({\bf x} - \hat{i}, t),
\label{eq:derivative-op}
\end{equation}
where the color indices are suppressed.
Equation~(\ref{eq:derivative-op}) defines a new smearing
distribution that is odd with respect to an inversion about point
${\bf x}$. Thus, smearing can contribute in a nontrivial way to
the behavior of the field with respect to inversion. This we call
smearing parity.
\subsection{Inversion, Parity and $\rho$-parity }
\label{subsec:ImproperPointGroup}

The improper point groups $O_h$ and $O^D_h$ consist of rotations
that leave the cube invariant together with the spatial inversion.
The parity transformation of a Dirac field involves
multiplication by the $\gamma_4$ Dirac matrix in addition to
spatial inversion as follows,
\begin{equation}
{\cal P} \tilde{q}({\bf x},t) {\cal P}^{-1} = \gamma_4
\tilde{q}(-{\bf x},t), \label{eq:parity}
\end{equation}
where ${\cal P}$ is the parity operator. Throughout this work we
employ the Dirac-Pauli representation for Dirac $\gamma$ matrices
for which $\gamma_4={\rm diag}[1,1,-1,-1]$. However, our results
may be used with any representation of the Dirac $\gamma$
matrices by applying the appropriate unitary transformation as
discussed in Appendix~\ref{app:DiracBasis}.

It is convenient to express the Dirac matrices as a direct product
of the form $SU(2)_\rho \ot SU(2)_s$ where the $ SU(2)$ components
are generated by the 2$\times$2 Pauli matrices for
 spin s and $\rho$-spin $\rho$~\cite{Gammel71,Kubis72}.
See Appendix~\ref{app:DiracBasis} for details of the construction.

Expressed in terms of the $SU(2)_\rho \ot SU(2)_s$ matrices,
$\gamma_4 = \rho_3\otimes \sigma_4$ where $\rho_3 = \rm{diag}[ 1,
-1]$ and $\sigma_4 = \rm{diag}[1, 1]$. Similarly, the Dirac
matrix $\sigma_{21}= {i \over 2} [\gamma_2,\gamma_1] = \rho_4
\otimes \sigma_3$ where $\rho_4 = \rm{diag}[ 1, 1]$ and $\sigma_3
= \rm{diag}[1, -1]$. With these conventions, a fermion field
satisfies
\begin{equation}
\gamma_4 \tilde{q} _{\mu}({\bf x},t) = (\rho_3 \otimes \sigma_4 )
\tilde{q} _{\mu}({\bf x},t) = \rho \tilde{q}_{\mu} ({\bf x},t),
\label{eq:rho_spin}
\end{equation}
and
\begin{equation}
\sigma_{21} \tilde{q} _{\mu}({\bf x},t) = (\rho_4 \otimes \sigma_3
) \tilde{q} _{\mu}({\bf x},t) = s \tilde{q}_{\mu} ({\bf x},t),
\label{eq:s_spin}
\end{equation}
where Table~\ref{table:Dirac_rho_s} provides the $\rho$ and $s$
values. Thus, the Dirac index $\mu = 1, 2, 3, 4$ is equivalent to
a two-dimensional superscript corresponding to $\rho$-spin ($\rho$
= +1, -1) and a two-dimensional subscript corresponding to spin
($s = +1, -1$), and the field may be written as $q^{\rho}_{s}({\bf
x},t)$. We refer to the $\rho$ value as $\rho$-parity because of
its role in the parity transformation of Eqs.~(\ref{eq:parity})
and (\ref{eq:rho_spin}).
\begin{table}[h]
\caption{Translation of the Dirac index $\mu$ to $\rho$- and
$s$-spin indices.  Index $\mu$ is expressed in the Dirac-Pauli
representation.} \label{table:Dirac_rho_s}
\renewcommand{\arraystretch}{1.0} 
\begin{center}
\begin{tabular}{|c|cc|}
\hline
Dirac index $\mu$ & ~~$\rho$~~ & ~~$s$~~ \\
 \hline
1 & $+$ & $+$ \\
2 & $+$ & $-$ \\
3 & $-$ & $+$ \\
4 & $-$ & $-$ \\
\hline
\end{tabular}
\end{center}
\end{table}

The parity transformation of a smeared quark field can differ from
that of an unsmeared field because the smearing parity enters.
This is most easily seen by using free fields for which the gauge
link variables are unity.  Then the smearing distribution does
not depend on the point ${\bf x}$ and reduces to the set of
coefficients $c({\bf y})$ that weight the fields at points ${\bf
y}$ away from the central point ${\bf x}$, i.e., the smeared
quark field is
\begin{equation}
q_{\mu}({\bf x},t) = \sum_{\bf y} c({\bf y})\tilde{q}_{\mu}({\bf
x} + {\bf y}, t)
 \end{equation}
 and the smearing parity is defined by
 \begin{equation}
 c(-{\bf y})= p c({\bf y}),
\end{equation}
where $p = +1$ or $-1$ for even or odd smearing parity,
respectively. The parity transformation of a smeared field is
\begin{eqnarray}
{\cal P} q_{\mu}({\bf x},t) {\cal P}^{-1} &=& \sum_{y} c({\bf y})
\gamma_4 \tilde{q}_\mu(-{\bf x}- {\bf y},t)
\nn \\
&=& \gamma_4  \sum_{\bf y}c(-{\bf y})\tilde{q}_{\mu}(-{\bf
x}+ {\bf y},t), \nn \\
&=& \gamma_4 p \sum_{y} c({\bf y})\tilde{q}_{\mu}(-{\bf
x}+ {\bf y},t), \nn \\
&=& \gamma_4 p q_{\mu}(-{\bf x},t),
 \label{eq:smeared_parity}
\end{eqnarray}
where the second line involves the relabeling ${\bf y}
\rightarrow - {\bf y}$ and the third line uses the symmetry of the
smearing distribution under inversion of ${\bf y}$. When the
gauge links are included so as to obtain a gauge covariant
smearing a similar result is obtained, which holds as an average
over gauge configurations.

 The parity transformation of a product of three smeared quark
fields is
\begin{eqnarray}
{\cal P} q^{\rho_1}_{s_1}({\bf x},t) q^{\rho_2}_{s_2}({\bf
x},t)q^{\rho_3}_{s_3}({\bf x},t) {\cal P}^{-1} = \nonumber
\\
\rho_1 \rho_2 \rho_3\,\,p_{1}p_{2}p_{3}\,\, q^{\rho_1}_{s_1}(-{\bf
x},t) q^{\rho_2}_{s_2}(-{\bf x},t)q^{\rho_3}_{s_3}(-{\bf x},t),
 \label{eq:3qparity}
\end{eqnarray}
where we have used the notation $q_{s}^{\rho} ({\bf x},t)$ in
place of $q_{\mu}({\bf x},t)$ and evaluated the $\gamma_4$
matrices using Eq.~(\ref{eq:rho_spin}) to obtain the product of
the three $\rho$-parities.  The product $\rho_1 \rho_2 \rho_3$ is
referred to simply as the $\rho$-parity of the operator and the
product $p_{1}p_{2}p_{3}$ is referred to as the smearing parity
of the operator.

The field operator at an arbitrary point ${\bf x}$ does not have
a definite parity.  However, in correlation functions projected to
zero total momentum, the ${\bf x}$ dependence is removed by a
translation following insertion of a complete set of intermediate
states, e.g.,
\begin{eqnarray}
C(t) &=& \sum_{\bf x}<0| B({\bf x},t) \bar{B}({\bf 0},0) |0> \nn
\\
&=& \sum_n <0| B({\bf 0},0)|n>e^{-M_nt}<n|\bar{B}({\bf 0},0)|0>.
\nn \\
 \label{eq:correlfn}
\end{eqnarray}
Thus the zero-momentum correlation function has baryon operators
only at point ${\bf x}= {\bf 0}$ where the operator has parity
given by the product of $\rho$-parity and smearing parity, i.e.,
$\rho_1 \rho_2 \rho_3\,\,p_{1}p_{2}p_{3}$ in
Eq.~(\ref{eq:3qparity}).  The parity of intermediate state $n$
must be the same in order to have a nonvanishing coupling.

Rotations of a quark field are generated by the Dirac matrices
$\sigma_{ij} = -\rho_4\otimes \sigma_k$ where indices i, j and k
are cyclic and take the values 1, 2 and 3.  Rotations are
diagonal in $\rho$-spin and thus give a linear combination of
fields with different $s$ labels but the same $\rho$-parity,
\begin{equation}
U(R) \o{q}_s^{\rho} ({\bf x},t) U^{\dag}(R) = \sum_{s'}
\o{q}_{s'}^{\rho} (R^{-1}{\bf x},t)T_{s's}(R),
\end{equation}
where $T_{s's}(R)$ is a representation matrix of rotation $R$.
This insight into the transformations of Dirac indices with
respect to rotations is the first reason that we find the
$\rho,s$ labels useful.

Note that a ``barred'' field transforms in the same way as a
quantum ``ket'' when the unitary quantum operator $U(R)$ is
applied, i.e.,
\begin{equation}
U(R) \l| s\r> = \sum_{s'}\l| s'\r> \l< s'\r|U(R) \l| s\r>
= \sum_{s'} \l| s'\r> T_{s's}(R).
\end{equation}
However,``unbarred'' fields also are required. Although they are
independent fields in the Euclidean theory, their transformations
are similar to those of quantum ``bra'' states,
\begin{equation}
U(R) q_s^{\rho} ({\bf x},t) U^{\dag}(R) = \sum_{s'}
q_{s'}^{\rho} (R^{-1}{\bf x},t)T^{\dag}_{ss'}(R).
\end{equation}
In this paper we state results generally in terms of ``barred''
fields in order to have a transparent connection between the
transformations of fields and those of the quantum states that
they create.  ``Unbarred'' operators generally involve the same
constructions except that coefficients or other operators
involved must be hermitian conjugated.

Operators that couple only to even parity intermediate states in
Eq.~(\ref{eq:correlfn}) are labeled with a subscript $g$ (for {\it
gerade}) and operators that couple only to odd parity states are
labeled with a subscript $u$ (for {\it ungerade}).  For
half-integer spins, the relevant IRs of $O^D_h$
are: $G_{1g}, G_{2g}, H_g, G_{1u}, G_{2u}, H_u$.
We will use these notations throughout this paper.

Because of the parity transformation of Eq.~(\ref{eq:3qparity}),
there are two independent ways to make baryon operators that
couple to states of a given parity in a zero-momentum correlation
function. Operators coupling to gerade states can be made either
with even smearing parity together with positive $\rho$-parity or
with odd smearing parity together with negative $\rho$-parity.
Similarly, there are two disjoint sets of operators that couple
to ungerade states: ones with odd smearing parity together with
positive $\rho$-parity or ones with even smearing parity together
with negative $\rho$-parity. These sets are not connected by
rotations because neither the smearing parity nor the
$\rho$-parity can be changed by a rotation. However, they are
connected by $\rho$-spin raising or lowering operations and in
our construction each operator that couples to a gerade state is
connected in this way with an operator that couples to an
ungerade state. This is the second reason that the $\rho,s$
labeling is useful. The $\rho , s$ labeling is used sparingly in
this paper but it is central to the method used in
Appendix~\ref{app:DiracSymmetry} to construct combinations of
Dirac indices that transform irreducibly.

Each baryon operator
carries a row label, $\lambda$, whose meaning
 depends upon the bases used for IRs.
The row label distinguishes between the $d_{\Lambda}$ members of
IR $\Lambda$.  If a representation contains more than one
occurrence of IR $\Lambda$, we say that there are multiple
embeddings of that IR.  A superscript, $k$, is used to
distinguish between the different embeddings. Therefore, a
generic baryon operator is denoted as $\o B^{\Lambda ,k
}_{\lambda} ({\bf x}, t)$, or in ``unbarred'' form as $B^{\Lambda
,k }_{\lambda} ({\bf x}, t)$, where the operator belongs to the
$k^{th}$ embedding of IR $\Lambda$ and row $\lambda$ of the
octahedral group.  Operators for different baryons are indicated
by the use of appropriate symbols, such as $\o N^{\Lambda ,k
}_{\lambda} ({\bf x}, t)$ (for isospin 1/2 operators), $\o \Delta
^{\Lambda ,k }_{\lambda} ({\bf x}, t)$ (for isospin 3/2
operators), $\o \Sigma^{\Lambda  ,k }_{\lambda} ({\bf x}, t)$, and
so on.

The correlations of operators belonging to different IRs or to
different rows of the same IR vanish:
\begin{eqnarray}
\sum_x \l<0\r| T B^{\Lambda  ,k }_\lambda({\bf x}, t) \o
B^{\Lambda',k' }_{\lambda'} (0)\l|0 \r> = C^{(\Lambda)}_{kk'}(t)
\delta_{\Lambda \Lambda'} \delta_{\lambda \lambda'}.
\label{eq:orthogonality}
\end{eqnarray}
The correlations of different embeddings of the same IR and row
are generally nonzero, providing sets of operators suitable for
constructing a correlation matrix $C^{(\Lambda)}_{kk'}(t)$.

\section{Quasi-local Baryonic Operators}
\label{sec:local}

Since the quark fields are Grassmann-valued and taken at a common
location ${\bf x}$, and the color indices are contracted with the
antisymmetric Levi-Civita tensor, our three-quark, quasi-local
baryon operators must be symmetric with respect to simultaneous
exchange of flavor and Dirac indices.
An operator that is symmetric in flavor labels ($\o \Delta, \o
\Omega$) must be symmetric also in Dirac indices, and an operator
that is mixed-antisymmetric in flavor labels ($\o N$) must be
mixed antisymmetric in Dirac indices, assuming that masses of the
up and down quarks are equal. An operator that is
mixed-antisymmetric in flavor labels and that has nonzero
strangeness ($\o \Lambda$) can have mixed-antisymmetric or
totally antisymmetric Dirac indices, and an operator that is
mixed-symmetric in flavor labels and that has nonzero strangeness
($\o \Sigma, \o \Xi$) can have mixed-symmetric or totally
symmetric Dirac indices. All possible symmetries of the Dirac
indices are encountered in the consideration of the different
baryons. In this section, we discuss the different baryons in turn
and develop tables of operators classified according to IRs of
$O_h^D$.

\subsection{Quasi-local Nucleon Operators}
\label{subsec:LocalNucleon}

Consider operators made from quasi-local quark fields for isospin
quantum numbers $I=1/2, I_z = 1/2$.  These operators correspond
to the $N^*$ family of baryons and they may be chosen to be
\begin{equation}
\o N^{+\rm (MA)}_{\mu_1\mu_2\mu_3}= \eps_{abc} {1\over \sqrt{2}}
\l( \bar{u}_{\mu_1}^{a} \bar{d}_{\mu_2}^{b} - \bar{d}_{\mu_1}^{a}
\bar{u}_{\mu_2}^{b} \r) \bar{u}_{\mu_3}^{c},
\label{eq:uudMA-local}
\end{equation}
where $u$ is an up quark and $d$ is a down quark. All (smeared) quark
fields are defined at spacetime point $({\bf x},t)$.
Equation~(\ref{eq:uudMA-local}) provides a proton
operator in the notation of the Particle Data Group~\cite{PDG}.
A neutron or $\o N^0$ operator can be obtained using the isospin
lowering operation.
\begin{center}
\begin{figure}[h]
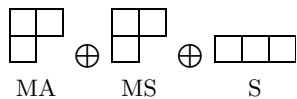

\begin{tabular}{ccccc}
\yng(2,1) & $\bigoplus$ & \yng(2,1) & $\bigoplus$ & \yng(3) \\
    MA    &             &      MS   &             &   S
\end{tabular}
\caption{ Young tableaux for three-quark $SU(2)_I$ irreducible
representations. The first tableau is antisymmetric in labels of
particles 1 and 2 (denoted MA for mixed-antisymmetric), while the
second tableau is symmetric in the labels of particles 1 and 2
(denoted MS for mixed-symmetric).  The third tableau is fully
symmetric (denoted S).}
\label{figure:IsoYoungTableaux}
\end{figure}
\end{center}
These two operators correspond to the mixed-antisymmetric Young
tableau for isospin in Fig.~\ref{figure:IsoYoungTableaux}. Each
$N^{\rm (MA)}_{\mu_1\mu_2\mu_3} ({\bf x},t)$ operator of
Eq.~(\ref{eq:uudMA-local}) is manifestly antisymmetric with
respect to the flavor interchange $u \leftrightarrow d$ applied
to the first two quark fields. This leads to the following
restrictions on Dirac indices,
\begin{eqnarray}
\o N^{\rm (MA)}_{\mu_1\mu_2\mu_3} + \o N^{\rm (MA)}_{\mu_2\mu_1\mu_3} &=& 0,
\label{eq:restriction1} \\
\o N^{\rm (MA)}_{\mu_1\mu_2\mu_3} + \o N^{\rm (MA)}_{\mu_2\mu_3\mu_1} +
\o N^{\rm (MA)}_{\mu_3\mu_1\mu_2} &=& 0. \label{eq:restriction2}
\end{eqnarray}

Some general considerations are stated most simply using the
Dirac indices. There are 4$^3 =$64 combinations of Dirac indices
for operators formed from three quark fields. They may be
classified by the four Young tableaux of
Fig.~\ref{figure:DiracYoungTableaux}, where each box is
understood to take the values $\mu = 1, 2, 3$ or $4$.
\renewcommand{\b}{$\bigoplus$}
\begin{center}
\begin{figure}[h]
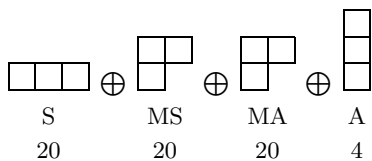

\begin{tabular}{ccccccc}
\yng(3) & \b & \yng(2,1) & \b & \yng(2,1) & \b & \yng(1,1,1) \\
   S    &    &    MS     &    &     MA    &    &      A      \\
  20    &    &    20     &    &     20    &    &      4
\end{tabular}
\caption{Young tableaux for three Dirac indices.}
\label{figure:DiracYoungTableaux}
\end{figure}
\end{center}
Standard rules for counting the dimensions of the tableaux show
that the totally symmetric tableau includes 20 operators, the
mixed-symmetric and mixed-antisymmetric tableaux each have 20
operators and the totally antisymmetric tableau has 4 operators,
thus accounting for all 64 possibilities. Groupings of Dirac
indices according to the symmetries of
Fig.~\ref{figure:DiracYoungTableaux} are useful. The fact that a
quasi-local baryonic operator must be symmetric with respect to
simultaneous exchange of flavor labels and Dirac indices
associates each baryon operator with one of the symmetries of
Dirac indices found in the tableaux of
Fig.~\ref{figure:DiracYoungTableaux}. All the combinations of
Dirac indices that correspond to each Young tableau are given
explicitly in Appendix~\ref{app:DiracSymmetry}. Three Dirac
spinors whose spin indices are written in accord with one Young
tableau in Fig.~\ref{figure:DiracYoungTableaux} form a closed set
under the group of rotations and parity transformations. These
group representations have been reduced to $G_{1g/u}$ and
$H_{g/u}$ IRs of Dirac indices by working with the $\rho,s$
labeling as discussed in Appendix~\ref{app:DiracSymmetry}.

Nucleon operators with MA isospin symmetry in
Eq.~(\ref{eq:uudMA-local}) must have the MA symmetry of Dirac
indices, corresponding to the third tableau of
Fig.~\ref{figure:DiracYoungTableaux}.
Table~\ref{table:LocalNucleon} gives explicit forms for the 20
quasi-local nucleon operators classified into IRs
$G_{1g},H_g,G_{1u},$ and $H_u$.  Here and in the remainder of
this paper we label the operators by the parity of intermediate
states to which they couple in a zero-momentum correlation
function, as in Eq.~(\ref{eq:correlfn}). Alternatively, one may
regard the operators in the tables as having been translated to
point ${\bf x} = {\bf 0}$, where they have definite parity as
seen from Eq.~(\ref{eq:smeared_parity}). Dirac indices in the
table come from Table~\ref{table:sixtyfour2} in
Appendix~\ref{app:DiracSymmetry}, but they have been simplified
using the relation in Eq.~(\ref{eq:restriction1}).
\renewcommand{\sp}{\!\! + \!\!}
  \newcommand{\sm}{\!\! - \!\!}
\begin{table}[h]
\caption{Quasi-local Nucleon operators. All operators have MA Dirac indices.}
\label{table:LocalNucleon}
\renewcommand{\arraystretch}{1.0} 
\begin{ruledtabular}
{\tabcolsep=0mm
\begin{tabular}{lc|lc}
$\o \Psi^{\Lambda,k}_{S,S_z}$ & $\o N_{\mu_1 \mu_2 \mu_3}$ &
$\o \Psi^{\Lambda,k}_{S,S_z}$ & $\o N_{\mu_1 \mu_2 \mu_3}$ \\
\hline
\hline
$\o \Psi^{G_{1g},1}_{{1\over 2},{1\over 2}}$&$\o N_{121}$ &
$\o \Psi^{G_{1u},1}_{{1\over 2},{1\over 2}}$&
${1\over\sqrt{3}}(\o N_{123}\sp\o N_{141}\sp\o N_{321})$\\
$\o \Psi^{G_{1g},1}_{{1\over 2},-{1\over 2}}$&$\o N_{122} $ &
$\o \Psi^{G_{1u},1}_{{1\over 2},-{1\over 2}}$&
${1\over\sqrt{3}}(\o N_{124}\sp\o N_{142}\sp\o N_{322})$\\
\hline
$\o \Psi^{G_{1g},2}_{{1\over 2},{1\over 2}}$&
${1\over\sqrt{3}}(\o N_{143}\sp\o N_{323}\sp\o N_{341})$&
$\o \Psi^{G_{1u},2}_{{1\over 2},{1\over 2}}$&$\o N_{343}$ \\
$\o \Psi^{G_{1g},2}_{{1\over 2},-{1\over 2}}$&
${1\over\sqrt{3}}(\o N_{144}\sp\o N_{324}\sp\o N_{342})$&
$\o \Psi^{G_{1u},2}_{{1\over 2},-{1\over 2}}$&$\o N_{344}$ \\
\hline
$\o \Psi^{G_{1g},3}_{{1\over 2},{1\over 2}}$&
${1\over\sqrt{3}}(\o N_{134}\sp\o N_{323}\sm\o N_{341})$&
$\o \Psi^{G_{1u},3}_{{1\over 2},{1\over 2}}$&
${1\over\sqrt{3}}(\!-\!\o N_{141}\sm\o N_{312}\sp\o N_{123})$ \\
$\o \Psi^{G_{1g},3}_{{1\over 2},-{1\over 2}}$&
${1\over\sqrt{3}}(\o N_{144}\sp\o N_{423}\sm\o N_{342})$&
$\o \Psi^{G_{1u},3}_{{1\over 2},-{1\over 2}}$&
${1\over\sqrt{3}}(\!-\!\o N_{322}\sm\o N_{241}\sp\o N_{124})$ \\
\hline
$\o \Psi^{H_g}_{{3\over 2},{3\over 2}}$&$\o N_{133}$ &
$\o \Psi^{H_u}_{{3\over 2},{3\over 2}}$&$ \o N_{131}$ \\
$\o \Psi^{H_g}_{{3\over 2},{1\over 2}}$&
${1\over\sqrt{3}}(\o N_{134}\sp\o N_{143}\sp\o N_{233})$ &
$\o \Psi^{H_u}_{{3\over 2},{1\over 2}}$&
${1\over\sqrt{3}}(\o N_{132}\sp\o N_{141}\sp\o N_{231})$ \\
$\o \Psi^{H_g}_{{3\over 2},-{1\over 2}}$&
${1\over\sqrt{3}}(\o N_{144}\sp\o N_{234}\sp\o N_{243})$ &
$\o \Psi^{H_u}_{{3\over 2},-{1\over 2}}$&
${1\over\sqrt{3}}(\o N_{142}\sp\o N_{232}\sp\o N_{241})$ \\
$\o \Psi^{H_g}_{{3\over 2},-{3\over 2}}$&$\o N_{244}$ &
$\o \Psi^{H_u}_{{3\over 2},-{3\over 2}}$&$\o N_{242}$ \\
\end{tabular}
}
\end{ruledtabular}
\end{table}
Because all the coefficients are real, ``unbarred'' operators are
obtained by replacing $\o N$ by $N$ in the same linear
combinations . The left column of the table shows 10 gerade
nucleon operators and the right one shows 10 ungerade operators.
For a given parity there are three sets of $G_1$ operators (three
embeddings of $G_1$) and one set of $H$ operators. Each $G_1$
IR contains two operators that transform amongst themselves
under rotations of the group and each $H$ IR contains four
operators that transform amongst themselves. Operators in each
IR are given spin projection labels, $S_z$, which are also
equivalent to ``row'' labels but more physically meaningful. In a
given embedding the operator with the largest $S_z$ is designated
``row~1'', the next largest $S_z$ is designated ``row~2'', and so
on. The notation $\o \Psi^{\Lambda,k}_{S,S_z}$ represents a
general quasi-local baryonic operator with spin $S$ and spin
projection $S_z$, transforming according to the $k$-th embedding
of IR $\Lambda$ of the group $ O^D_h$.

Spin-raising and spin-lowering operators for a Dirac spinor are
\begin{equation}
s^\pm = {1\over 2}\l(
\begin{array}{cc}
\sigma_1 \pm i\sigma_2 & 0 \\
0 & \sigma_1 \pm i\sigma_2
\end{array}
\r),
\label{eq:s-}
\end{equation}
in the Dirac-Pauli representation. For a three-quark state, the
spin raising or lowering operator is a sum of three terms, for
example, $S^-= \sum_{j=1}^3 s_j^\pm$, where $s^\pm_j$ acts on the
$j$-th quark. The same operations carry over to the ``barred''
field operators of Table~\ref{table:LocalNucleon}. Different rows
in the same embedding of an IR are related to one another by
spin raising and lowering operations. For example, the
transformation of the first $G_1$ embedding of
Table~\ref{table:LocalNucleon} proceeds schematically as follows,
\begin{eqnarray}
\hspace{-3mm}
S^- \o \Psi^{G_{1g},1}_{1/2, 1/2}=S^- \o N _{121}
\hspace{-2mm} &=& \hspace{-2mm} S^- \o N^{+++}_{+-+}
\nn \\*
&=& \hspace{-2mm}
\o N^{+++}_{+--} = \o N_{122} = \o
\Psi^{G_{1g},1}_{1/2,-1/2},~ \label{eq:nucleon_s-}
\end{eqnarray}
where the notation $\o N^{\rho_1 \rho_2 \rho_3}_{s_1 s_2 s_3}$ is
used in the intermediate steps. Note that a spin-lowering
operation on the second quark in Eq.~(\ref{eq:nucleon_s-})
vanishes because it has spin down and spin-lowering of the first
quark also vanishes because $s_1^- \o N^{+++}_{+-+}=\o
N^{+++}_{--+}=0$ by Eq.~(\ref{eq:restriction1}). Spin raising and
lowering operations can be applied repeatedly and the following
relation holds,
\newcommand{\arS}{\longrightarrow \hspace{-6mm}
                 {\raise 0.5mm \hbox{$ ^{S^-}$}} \hspace{3mm}}
\newcommand{\alS}{\longleftarrow \hspace{-6mm}
                 {\lower 1.1mm \hbox{$ _{S^+}$}} \hspace{3mm}}
\newcommand{\alrS}{                                      \hspace{ 2.0mm}
                  {\lower 0.7mm \hbox{$\longleftarrow$}} \hspace{-6.5mm}
                  {\raise 0.7mm \hbox{$\longrightarrow$}}\hspace{-4.6mm}
          {\lower 1.8mm \hbox{$ _{S^+}$}}        \hspace{-4.4mm}
          {\raise 1mm   \hbox{$ ^{S^-}$}}        \hspace{ 3.0mm}}
\begin{eqnarray}
0 \alS \o \Psi^{G_{1g/u},k}_{1/2,
1/2}\alrS \o \Psi^{G_{1g/u},k}_{1/2,-1/2}  \arS 0, \nn \\
0 \alS \o \Psi^{H_{g/u},k}_{3/2,3/2} \alrS \o
\Psi^{H_{g/u},k}_{3/2, 1/2}  \alrS \o \Psi^{H_{g/u},k}_{3/2,-1/2}
\nn \\* \alrS \o \Psi^{H_{g/u},k}_{3/2,-3/2} \arS 0 .
\label{eq:s_rule}
\end{eqnarray}

A gerade operator in a row of Table~\ref{table:LocalNucleon} and
the ungerade one in the same row are related to each other by
$\rho$-spin raising and lowering operations. For Dirac spinors,
the operators are $\rho^\pm = \sum_{j=1}^3 \rho_j^\pm$,
$\rho^\pm_j = (1/2) (\rho^1_j \pm i \rho^2_j)$, where $j$
specifies the first, second, or the third quark.
An example is
\begin{eqnarray}
\rho^- \o \Psi^{G_{1g},1}_{1/2, 1/2} &=&
\rho^- \o N_{121}=\rho^- \o N^{+++}_{+-+},\nn \\
&\rightarrow& {1\over\sqrt{3}}
\l( \o N^{++-}_{+-+} + \o N^{+-+}_{+-+}+ \o N^{-++}_{+-+} \r) \nn \\
 &=& {1\over\sqrt{3}} \l( \o N_{123} + \o N_{141} + \o N_{321} \r) \nn \\
 &=& \o \Psi^{G_{1u},1}_{1/2,1/2},
\label{eq:rho_rule}
\end{eqnarray}
where an appropriate normalization is included in the resultant
operator. Note that $\rho$-spin raising and lowering operations
change the $\rho$-parity of one quark and thus change the product
$\rho_1\rho_2\rho_3$ which is the $\rho$-parity of the operator.
However, they preserve the $s$ labels and leave the transformation
properties under rotations unchanged.
\newcommand{\arm}{\longrightarrow \hspace{-6mm}
                {\raise 0.6mm \hbox{$ ^{\rho^-}$}} \hspace{3mm}}
\newcommand{\arp}{\longrightarrow \hspace{-6mm}
                {\raise 0.6mm \hbox{$ ^{\rho^+}$}} \hspace{3mm}}
\newcommand{\al}{\longleftarrow \hspace{-6mm}
                {\lower 0.6mm \hbox{$ _{\rho^+}$}} \hspace{3mm}}
\newcommand{\alr}{                                      \hspace{ 2.0mm}
                 {\lower 0.7mm \hbox{$\longleftarrow$}} \hspace{-6.5mm}
                 {\raise 0.7mm \hbox{$\longrightarrow$}}\hspace{-4.3mm}
         {\lower 1.6mm \hbox{$ _{\rho^+}$}}     \hspace{-3.7mm}
         {\raise 1.5mm \hbox{$ ^{\rho^-}$}}     \hspace{ 3.0mm}}

Mixed-symmetric isospin operators with $I=1/2, I_z = 1/2$
may also be defined by
\begin{equation}
\o N^{+\rm (MS)}_{\mu_1 \mu_2 \mu_3} = \eps_{abc}
 \frac{1}{\sqrt{6}} ( 2 \bar{u}_{\mu_1}^a \bar{u}_{\mu_2}^b \bar{d}_{\mu_3}^c -
 \bar{u}_{\mu_1}^a \bar{d}_{\mu_2}^b \bar{u}_{\mu_3}^c -
 \bar{d}_{\mu_1}^a \bar{u}_{\mu_2}^b \bar{u}_{\mu_3}^c ) .
\label{eq:uudMS-local}
\end{equation}
However, for quasi-local operators they can be rewritten in terms
of the MA isospin operators defined in
Eq.~(\ref{eq:uudMA-local}) as follows,
\begin{equation}
\o N^{\rm (MS)}_{\mu_1\mu_2\mu_3}= \o N^{\rm
(MA)}_{\mu_3\mu_2\mu_1}+\o N^{\rm (MA)}_{\mu_1\mu_3\mu_2},
\label{eq:local_NMA_NMS}
\end{equation}
showing that the quasi-local, MS isospin operators
are not linearly independent of quasi-local MA
operators. It is sufficient to consider only the
MA operators of Eq.~(\ref{eq:uudMA-local}) in
order to construct a complete, linearly independent set of isospin
1/2, quasi-local operators.

Relations of the operators presented in
Table~\ref{table:LocalNucleon} to operators that are
commonly used in the literature for a nucleon are
given in Appendix~\ref{app:chi1_chi2}.

\subsection{Quasi-local $\Delta$ and $\Omega$ Operators}
\label{subsec:LocalDelta}

The isospin of a $\Delta$ baryon is 3/2 and there are four
different operators corresponding to isospin projections
$I_z=3/2,1/2,-1/2$, and $-3/2$:
\newcommand{\D}{\o \Delta}
\begin{eqnarray}
\D^{\, ++}_{\mu_1 \mu_2 \mu_3} &=& \eps_{a b c} \:
\bar{u}^{a}_{\mu_1}\bar{u}^{b}_{\mu_2}\bar{u}^{c}_{\mu_3}, \nn \\
\D^{\, +}_{\mu_1 \mu_2 \mu_3} &=& \frac{\eps_{a b c}}{\sqrt{3}}
\l( \bar{u}^{a}_{\mu_1}\bar{u}^{b}_{\mu_2}\bar{d}^{c}_{\mu_3} +
\bar{u}^{a}_{\mu_1}\bar{d}^{b}_{\mu_2}\bar{u}^{c}_{\mu_3} +
\bar{d}^{a}_{\mu_1}\bar{u}^{b}_{\mu_2}\bar{u}^{c}_{\mu_3} \r) , \nn \\
\D^{\, 0}_{\mu_1 \mu_2 \mu_3} &=& \frac{\eps_{a b c}}{\sqrt{3}}
\l( \bar{u}^{a}_{\mu_1}\bar{d}^{b}_{\mu_2}\bar{d}^{c}_{\mu_3} +
\bar{d}^{a}_{\mu_1}\bar{u}^{b}_{\mu_2}\bar{d}^{c}_{\mu_3} +
\bar{d}^{a}_{\mu_1}\bar{d}^{b}_{\mu_2}\bar{u}^{c}_{\mu_3} \r) , \nn \\
\D^{\, -}_{\mu_1 \mu_2 \mu_3} &=& \eps_{a b c} \:
\bar{d}^{a}_{\mu_1}\bar{d}^{b}_{\mu_2}\bar{d}^{c}_{\mu_3}, \nn \\
\label{eq:LocalDelta}
\end{eqnarray}
where all fields are defined at spacetime point $({\bf x}, t)$.
Because of the totally symmetric flavors, the $\Delta$ baryon
operators must have totally symmetric combinations of Dirac
indices. According to Table~\ref{table:sixtyfour1} in
Appendix~\ref{app:DiracSymmetry} there are 20 combinations of
totally symmetric Dirac indices.  In a color-singlet three-quark
operator, the quark fields may be commuted with one another with
no change of sign.  This allows Dirac indices to be rearranged to
a standard order in which they do not decrease from left to
right, producing the 20 irreducible operators that are given in
Table~\ref{table:LocalDelta}.
%
\begin{table}[h]
\caption{Quasi-local $\Delta$ operators.
All operators have Dirac indices in S.}
\label{table:LocalDelta}
\renewcommand{\arraystretch}{1.1} 
\begin{ruledtabular}
\begin{tabular}{lc|lc}
$\o \Psi^{\Lambda,k}_{S,S_z}$ & ~~~~~$\D_{\mu_1 \mu_2 \mu_3}$~~~~~~ &
$\o \Psi^{\Lambda,k}_{S,S_z}$ & $\D_{\mu_1 \mu_2 \mu_3}$ \\
\hline
\hline
$\o \Psi^{G_{1g},1}_{{1\over 2}, {1\over 2}}$&$\D_{134} - \D_{233}$ &
$\o \Psi^{G_{1u},1}_{{1\over 2}, {1\over 2}}$&$\D_{114} - \D_{123}$ \\
$\o \Psi^{G_{1g},1}_{{1\over 2},-{1\over 2}}$&$\D_{144} - \D_{234}$ &
$\o \Psi^{G_{1u},1}_{{1\over 2},-{1\over 2}}$&$\D_{124} - \D_{223}$ \\
\hline
$\o \Psi^{H_g,1}_{{3\over 2}, {3\over 2}}$&$\D_{111}$ &
$\o \Psi^{H_u,1}_{{3\over 2}, {3\over 2}}$&$\sqrt{3}~\D_{113}$ \\
$\o \Psi^{H_g,1}_{{3\over 2}, {1\over 2}}$&$\sqrt{3}~\D_{112}$ &
$\o \Psi^{H_u,1}_{{3\over 2}, {1\over 2}}$&$\D_{114}+2\D_{123}$ \\
$\o \Psi^{H_g,1}_{{3\over 2},-{1\over 2}}$&$\sqrt{3}~\D_{122}$ &
$\o \Psi^{H_u,1}_{{3\over 2},-{1\over 2}}$&$2\D_{124}+\D_{223}$ \\
$\o \Psi^{H_g,1}_{{3\over 2},-{3\over 2}}$&$\D_{222}$ &
$\o \Psi^{H_u,1}_{{3\over 2},-{3\over 2}}$&$\sqrt{3}~\D_{224}$ \\
\hline
$\o \Psi^{H_g,2}_{{3\over 2}, {3\over 2}}$&$\sqrt{3}~\D_{133}$ &
$\o \Psi^{H_u,2}_{{3\over 2}, {3\over 2}}$&$\D_{333}$ \\
$\o \Psi^{H_g,2}_{{3\over 2}, {1\over 2}}$&$2\D_{134}+\D_{233}$ &
$\o \Psi^{H_u,2}_{{3\over 2}, {1\over 2}}$&$\sqrt{3}~\D_{334}$ \\
$\o \Psi^{H_g,2}_{{3\over 2},-{1\over 2}}$&$\D_{144}+2\D_{234}$ &
$\o \Psi^{H_u,2}_{{3\over 2},-{1\over 2}}$&$\sqrt{3}~\D_{344}$ \\
$\o \Psi^{H_g,2}_{{3\over 2},-{3\over 2}}$&$\sqrt{3}~\D_{244}$ &
$\o \Psi^{H_u,2}_{{3\over 2},-{3\over 2}}$&$\D_{444}$ \\
\end{tabular}
\end{ruledtabular}
\end{table}
 For each parity, two
embeddings of the $H$ IR occur, while there is one embedding
of the $G_1$ IR.  Table~\ref{table:LocalDelta} holds for any
$I_z$ value.  Spin-raising and lowering operations as in
Eq.~(\ref{eq:s_rule}) and $\rho$-spin raising and lowering
operations as in Eq.~(\ref{eq:rho_rule}) can be applied
 to relate operators in different
rows or operators in different columns of
Table~\ref{table:LocalDelta}.

The $\Omega$ baryons are composed of three strange quarks,
\begin{equation}
\o \Omega^{\, -}_{\mu_1 \mu_2 \mu_3}({\bf x}, t) = \eps_{a b c} \:
\bar{s}^{a}_{\mu_1}({\bf x}, t)\bar{s}^{b}_{\mu_2}({\bf x},
t)\bar{s}^{c}_{\mu_3}({\bf x}, t). \label{eq:LocalOmega}
\end{equation}
The quark flavors are clearly totally symmetric so only the
totally symmetric Dirac indices are allowed.  Therefore,
Table~\ref{table:LocalDelta} can be used for an $\Omega$ baryon.
In summary the $\D$ symbol in Table~\ref{table:LocalDelta} may be
replaced with any of $\{ \D^{\,++},\D^{\,+},\o \Delta ^{\,
0},\D^{\, -},\o \Omega^{\,-} \}$.

\subsection{Quasi-local $\Lambda$ Baryon Operators}
\label{subsec:LocalLambda}

The $\Lambda$ baryons have isospin zero and strangeness $-1$.
Appropriate quasi-local $\Lambda$ baryon operators have the form,
\begin{equation}
\o \Lambda^0_{\mu_1 \mu_2 \mu_3} = \eps_{a b c} {1\over\sqrt{2}}
\l( \bar{u}^{a}_{\mu_1} \bar{d}^{b}_{\mu_2} - \bar{d}^{a}_{\mu_1}
u^{b}_{\mu_2} \r) \bar{s}^{c}_{\mu_3} , \label{eq:LocalLambda}
\end{equation}
where spacetime arguments $({\bf x}, t)$ are omitted from each
quark field.  The $\Lambda$ baryon operator has a pair of up and
down quarks in the isospin zero state, which is the same as the
mixed-antisymmetric nucleon operator. Because the operator in
Eq.~(\ref{eq:LocalLambda}) satisfies the relation
\begin{equation}
\o \Lambda^0_{\mu_1 \mu_2 \mu_3} + \o \Lambda^0_{\mu_2 \mu_1
\mu_3} = 0 , \label{eq:restriction_lambda}
\end{equation}
it is antisymmetric with respect to exchange of $\mu_1$ and
$\mu_2$ indices.  Allowed symmetries of Dirac indices for the
quasi-local $\Lambda$ baryon operator are mixed-antisymmetric and
totally antisymmetric. The difference from the quasi-local nucleon
operator is that the $\Lambda$ baryon operator is allowed to have
totally antisymmetric Dirac indices, because the strange quark
removes the restriction of Eq.~(\ref{eq:restriction2}).
Table~\ref{table:LocalLambda} gives all quasi-local $\Lambda$
baryon operators.
\renewcommand{\L}{\o \Lambda}
\begin{table}[h]
\caption{Quasi-local $\Lambda$ baryon operators.}
\label{table:LocalLambda}
\begin{ruledtabular}
{\tabcolsep=0mm
\begin{tabular}{lc|lc}
$\o \Psi^{\Lambda,k}_{S,S_z}$& $\L_{\mu_1 \mu_2 \mu_3}$ &
$\o \Psi^{\Lambda,k}_{S,S_z}$& $\L_{\mu_1 \mu_2 \mu_3}$ \\
\hline
\hline
$\o \Psi^{G_{1g},1}_{{1\over 2},{1\over 2}}$ &$\L_{121}$ &
$\o \Psi^{G_{1u},1}_{{1\over 2},{1\over 2}}$ &$\L_{123}+\L_{141}+\L_{321}$\\
$\o \Psi^{G_{1g},1}_{{1\over 2},-{1\over 2}}$ &$\L_{122} $ &
$\o \Psi^{G_{1u},1}_{{1\over 2},-{1\over 2}}$ &$\L_{124}+\L_{142}+\L_{322}$\\
\hline
$\o \Psi^{G_{1g},2}_{{1\over 2},{1\over 2}}$&$\L_{143}+\L_{323}+\L_{341}$ &
$\o \Psi^{G_{1u},1}_{{1\over 2},{1\over 2}}$ &$\L_{343}$ \\
$\o \Psi^{G_{1g},2}_{{1\over 2},-{1\over 2}}$&$\L_{144}+\L_{324}+\L_{342}$&
$\o \Psi^{G_{1u},1}_{{1\over 2},-{1\over 2}}$ &$\L_{344}$ \\
\hline
$\o \Psi^{G_{1g},3}_{{1\over 2},{1\over 2}}$&$\L_{134}+\L_{323}-\L_{341}$&
$\o \Psi^{G_{1u},1}_{{1\over 2},{1\over 2}}$ &$-\L_{141}-\L_{312}+\L_{123}$\\
$\o \Psi^{G_{1g},3}_{{1\over 2},-{1\over 2}}$&$\L_{144}+\L_{423}-\L_{342}$&
$\o \Psi^{G_{1u},1}_{{1\over 2},-{1\over 2}}$ &$-\L_{322}-\L_{241}+\L_{124}$\\
\hline
$\o \Psi^{G_{1g},4}_{{1\over 2},{1\over 2}}$&
$\sqrt{2\over 3} (\L_{134}\sp\L_{341}\sp\L_{413})$&
$\o \Psi^{G_{1u},4}_{{1\over 2},{1\over 2}}$ &
$-\sqrt{2\over 3}(\L_{123}\sp\L_{231}\sp\L_{312})$ \\
$\o \Psi^{G_{1g},4}_{{1\over 2},-{1\over 2}}$&
$\sqrt{2\over 3} (\L_{234}\sp\L_{342}\sp\L_{423})$&
$\o \Psi^{G_{1u},4}_{{1\over 2},-{1\over 2}}$ &
$-\sqrt{2\over 3}(\L_{124}\sp\L_{241}\sp\L_{412})$ \\
\hline
$\o \Psi^{H_g}_{{3\over 2},{3\over 2}}$ &$\L_{133}$ &
$\o \Psi^{H_u}_{{3\over 2},{3\over 2}}$ &$ \L_{131}$ \\
$\o \Psi^{H_g}_{{3\over 2},{1\over 2}}$ &$\L_{134}+\L_{143}+\L_{233}$ &
$\o \Psi^{H_u}_{{3\over 2},{1\over 2}}$ &$\L_{132}+\L_{141}+\L_{231}$ \\
$\o \Psi^{H_g}_{{3\over 2},-{1\over 2}}$ &$\L_{144}+\L_{234}+\L_{243}$ &
$\o \Psi^{H_u}_{{3\over 2},-{1\over 2}}$ &$\L_{142}+\L_{232}+\L_{241}$ \\
$\o \Psi^{H_g}_{{3\over 2},-{3\over 2}}$ &$\L_{244}$ &
$\o \Psi^{H_u}_{{3\over 2},-{3\over 2}}$ & $\L_{242}$ \\
\end{tabular}
}
\end{ruledtabular}
\end{table}
Twelve positive-parity operators are given in the left half of the
table and twelve negative-parity operators are given in the right
half. Only four combinations of Dirac indices are totally
antisymmetric under exchange, and they belong to $G_1$ IRs.
Together with the three embeddings of $G_1$ that come from
mixed-antisymmetric combinations of Dirac indices, this provides a
total of four embeddings of $G_1$ in each parity plus one
embedding of the $H$ IR for quasi-local $\Lambda$ baryon
operators.

Irreducible basis operators for $\Lambda_c$ and $\Lambda_b$ baryons are
exactly the same except that the third quark is replaced by a charm or
bottom quark.

\subsection{Quasi-local $\Sigma$ and $\Xi$ Operators}
\label{subsec:LocalSigma}

A $\Sigma$ baryon has two light quarks forming an isospin triplet
combination and a strange quark.  Suitable $\Sigma$ operators are
defined such that the first two Dirac indices refer to the light
quarks,
\begin{eqnarray}
\o \Sigma^+_{\mu_1\mu_2\mu_3}&=&\eps_{abc}
\:\bar{u}^a_{\mu_1} \bar{u}^b_{\mu_2} \bar{s}^c_{\mu_3},\nn\\
\o \Sigma^0_{\mu_1\mu_2\mu_3}&=&\eps_{abc} {1\over\sqrt{2}}
\l(\bar{u}^a_{\mu_1} \bar{d}^b_{\mu_2}
+ \bar{d}^a_{\mu_1} \bar{u}^b_{\mu_2} \r) \bar{s}^c_{\mu_3}, \nn \\
\o \Sigma^-_{\mu_1\mu_2\mu_3}&=&\eps_{abc}\:\bar{d}^a_{\mu_1}
\bar{d}^b_{\mu_2} \bar{s}^c_{\mu_3}. \label{eq:LocalSigma}
\end{eqnarray}
Such operators satisfy the relation,
\begin{equation}
\o \Sigma_{\mu_1 \mu_2 \mu_3} - \o \Sigma_{\mu_2 \mu_1 \mu_3} =
0, \label{eq:restriction_sigma}
\end{equation}
showing that the Dirac indices must be totally symmetric or
mixed-symmetric.

\renewcommand{\S}{\o \Sigma}
\begin{table*}
\begin{center}
\caption{Quasi-local $\Sigma$ operators.}
\label{table:LocalSigma}
\renewcommand{\arraystretch}{1.2} 
\begin{tabular}{|lc|lc|}
\hline
$\o \Psi^{\Lambda,k}_{S,S_z}$ & $\S_{\mu_1 \mu_2 \mu_3}$ notation &
$\o \Psi^{\Lambda,k}_{S,S_z}$ & $\S_{\mu_1 \mu_2 \mu_3}$ notation \\
\hline \hline $\o \Psi^{G_{1g},1}_{{1\over2},{1\over2}}$&
     ${\sqrt{2}\over 3}[-\S_{332}-2\S_{233}+\S_{341}+\S_{413}+\S_{134}]$&
$\o \Psi^{G_{1u},1}_{{1\over2}, {1\over2}}$&
     ${\sqrt{2}\over3}[\S_{114}+2\S_{141}-\S_{123}-\S_{312}-\S_{231}]$\\
$\o \Psi^{G_{1g},1}_{{1\over2},-{1\over2}}$&
     ${\sqrt{2}\over3}[2\S_{144} +\S_{441}-\S_{234}-\S_{342}-\S_{423}]$&
$\o \Psi^{G_{1u},1}_{{1\over2},-{1\over2}}$&
     ${\sqrt{2}\over3}[-\S_{223}-2\S_{232}+\S_{124}+\S_{241}+\S_{412}]$\\
\hline
$\o \Psi^{G_{1g},2}_{{1\over2},{1\over2}}$&
     $\sqrt{2\over3}[\S_{112}-\S_{121}]$ &
$\o \Psi^{G_{1u},2}_{{1\over2}, {1\over2}}$&
     ${\sqrt{2}\over3}[\S_{114}+2\S_{312}-\S_{123}-\S_{141}-\S_{231}]$ \\
$\o \Psi^{G_{1g},2}_{{1\over2},-{1\over2}}$&
     $\sqrt{2\over3}[-\S_{221}+\S_{122}]$ &
$\o \Psi^{G_{1u},2}_{{1\over2},-{1\over2}}$&
     ${\sqrt{2}\over3}[-\S_{223}-2\S_{241}+\S_{124}+\S_{232}+\S_{412}]$ \\
\hline
$\o \Psi^{G_{1g},3}_{{1\over2},{1\over2}}$&
     ${\sqrt{2}\over 3}[\S_{332}+2\S_{134}-\S_{341}-\S_{233}-\S_{413}]$ &
$\o \Psi^{G_{1u},3}_{{1\over2}, {1\over2}}$&
     $\sqrt{2\over3}[\S_{334}-\S_{343}]$ \\
$\o \Psi^{G_{1g},3}_{{1\over2},-{1\over2}}$&
     ${\sqrt{2}\over 3}[-\S_{441}-2\S_{423}+\S_{342}+\S_{144}+\S_{234}]$ &
$\o \Psi^{G_{1u},3}_{{1\over2},-{1\over2}}$&
     $\sqrt{2\over3}[-\S_{443}+\S_{344}]$ \\
\hline
$\o \Psi^{G_{1g},4}_{{1\over2},{1\over2}}$&
     ${\sqrt{2}\over 3}[-\S_{332}-2\S_{413}+\S_{233}+\S_{134}+\S_{341}]$ &
$\o \Psi^{G_{1u},4}_{{1\over2}, {1\over2}}$&
     ${\sqrt{2}\over3}[\S_{114}+2\S_{231}-\S_{141}-\S_{312}-\S_{123}]$ \\
$\o \Psi^{G_{1g},4}_{{1\over2},-{1\over2}}$&
     ${\sqrt{2}\over 3}[\S_{441}+2\S_{234}-\S_{144}-\S_{423}-\S_{342}]$ &
$\o \Psi^{G_{1u},4}_{{1\over2},-{1\over2}}$&
     ${\sqrt{2}\over3}[-\S_{223}-2\S_{412}+\S_{232}+\S_{241}+\S_{124}]$ \\
\hline
$\o \Psi^{H_g,1}_{{3\over2}, {3\over2}}$& $\S_{111}$ &
$\o \Psi^{H_u,1}_{{3\over2}, {3\over2}}$&
${1\over\sqrt{3}}[\S_{113}+2\S_{131}]$ \\
$\o \Psi^{H_g,1}_{{3\over2}, {1\over2}}$&
${1\over\sqrt{3}}[\S_{112}+2\S_{121}]$ &
$\o \Psi^{H_u,1}_{{3\over2}, {1\over2}}$&
     ${1\over3}[2\S_{141}+\S_{114}+2\S_{312}+2\S_{123}+2\S_{231}]$\\
$\o \Psi^{H_g,1}_{{3\over2},-{1\over2}}$&
${1\over\sqrt{3}}[2\S_{122}+\S_{221}]$ &
$\o \Psi^{H_u,1}_{{3\over2},-{1\over2}}$&
     ${1\over3}[2\S_{232}+\S_{223}+2\S_{412}+2\S_{124}+2\S_{241}]$\\
$\o \Psi^{H_g,1}_{{3\over2},-{3\over2}}$& $\S_{222}$ &
$\o \Psi^{H_u,1}_{{3\over2},-{3\over2}}$&
${1\over\sqrt{3}}[\S_{224}+2\S_{242}]$ \\
\hline
$\o \Psi^{H_g,2}_{{3\over2}, {3\over2}}$&
${1\over\sqrt{3}}[2\S_{133}+\S_{331}]$ &
$\o \Psi^{H_u,2}_{{3\over2}, {3\over2}}$& $\S_{333}$ \\
$\o \Psi^{H_g,2}_{{3\over2}, {1\over2}}$&
     ${1\over3}[2\S_{233}+\S_{332}+2\S_{134}+2\S_{341}+2\S_{413}]$&
$\o \Psi^{H_u,2}_{{3\over2}, {1\over2}}$&
${1\over\sqrt{3}}[\S_{334}+2\S_{343}]$ \\
$\o \Psi^{H_g,2}_{{3\over2},-{1\over2}}$&
     ${1\over3}[2\S_{144}+\S_{441}+2\S_{234}+2\S_{342}+2\S_{423}]$&
$\o \Psi^{H_u,2}_{{3\over2},-{1\over2}}$&
${1\over\sqrt{3}}[2\S_{344}+\S_{443}]$ \\
$\o \Psi^{H_g,2}_{{3\over2},-{3\over2}}$&$(1/\sqrt{3})[2\S_{244}+\S_{442}]$&
$\o \Psi^{H_u,2}_{{3\over2},-{3\over2}}$& $\S_{444}$ \\
\hline
$\o \Psi^{H_g,3}_{{3\over2}, {3\over2}}$&
$\sqrt{2\over3}[-\S_{331}+\S_{133}]$ &
$\o \Psi^{H_u,3}_{{3\over2}, {3\over2}}$&
$\sqrt{2\over3}[\S_{113}-\S_{131}]$ \\
$\o \Psi^{H_g,3}_{{3\over2}, {1\over2}}$&
     ${\sqrt{2}\over3}[-\S_{332}-2\S_{341}+\S_{134}+\S_{233}+\S_{413}]$&
$\o \Psi^{H_u,3}_{{3\over2}, {1\over2}}$&
     ${\sqrt{2}\over3}[\S_{114}+2\S_{123}-\S_{312}-\S_{141}-\S_{231}]$\\
$\o \Psi^{H_g,3}_{{3\over2},-{1\over2}}$&
     ${\sqrt{2}\over3}[-2\S_{342}-\S_{441}+\S_{234}+\S_{144}+\S_{423}]$&
$\o \Psi^{H_u,3}_{{3\over2},-{1\over2}}$&
     ${\sqrt{2}\over3}[2\S_{124}+\S_{223}-\S_{412}-\S_{232}-\S_{241}]$\\
$\o \Psi^{H_g,3}_{{3\over2},-{3\over2}}$&
$\sqrt{2\over3}[-\S_{442}+\S_{244}]$ &
$\o \Psi^{H_u,3}_{{3\over2},-{3\over2}}$&
$\sqrt{2\over3}[\S_{224}-\S_{242}]$ \\
\hline
\end{tabular}
\end{center}
\end{table*}

A $\Xi$ baryon has two strange quarks and one light quark forming
an isospin doublet,
\begin{eqnarray}
\o \Xi^0 &=& \eps_{abc} \:
\bar{s}^a_{\mu_1} \bar{s}^b_{\mu_2} \bar{u}^c_{\mu_3}, \nn \\
\o \Xi^- &=& \eps_{abc} \: \bar{s}^a_{\mu_1} \bar{s}^b_{\mu_2}
\bar{d}^c_{\mu_3}.
\end{eqnarray}
Again the operators are symmetric under the exchange of $\mu_1$
and $\mu_2$. Thus, the allowed combinations of Dirac indices are
the same as for the quasi-local $\Sigma$ baryon operators.
Table~\ref{table:LocalSigma} presents all operators with
symmetric and mixed-symmetric Dirac indices. Note that there are
20 operators for totally symmetric Dirac indices (as in the
quasi-local $\Delta$ operators) and 20 operators for
mixed-symmetric Dirac indices, giving a total of 40 operators for
$\Sigma$ or $\Xi$ baryons.  Four $G_1$ embeddings and three $H$
embeddings occur in each parity.  In Table~\ref{table:LocalSigma}
the symbol $\o \Sigma$ may be replaced by any of $\{ \o
\Sigma^+,\o \Sigma^0,\o \Sigma^-,\o \Xi^0,\o \Xi^- \}$.

\section{Nonlocal Baryonic Operators}
\label{sec:nonlocal}

In this section we discuss how to construct baryon operators that
create states whose wave functions have angular or radial
excitation. Orbital angular momentum or radial excitation is
expected to be of particular interest for operators that couple
to excited baryons.

In Section~\ref{sec:local}, all possible symmetries of the
Dirac indices of three quarks were encountered.  When nonlocal
operators are constructed, we can build upon the quasi-local
operators already found by adding a nontrivial spatial structure.
This basically amounts to allowing different smearings of the
quark fields.

Nonlocal operators are constructed by displacing at least
one quark from the others.
The set of displacements is first
arranged to belong to the basis of IRs of the octahedral group.
Then there arises the issue of combining the IRs of spatial
distributions of displacements with IRs
of the Dirac indices that have been developed for quasi-local
operators.  With respect to the octahedral group, the spatial and
spin IRs transform as direct products. Using Clebsch-Gordan
coefficients, we form linear combinations of the direct products
so as to obtain nonlocal operators that transform as overall
IRs of the group.


\subsection{Displaced quark fields and IRs of $O$ }
\label{sec:displacement}

Relative displacement of quarks requires insertion of a
path-dependent gauge link in order to maintain gauge invariance.
The simplest such displaced three-quark operator would be of the
form,
\begin{equation}
  \o{b}^{(i) f_1 f_2 f_3}_{\mu_1 \mu_2 \mu_3} ({\bf x}) =
  \eps_{a b c} \:
  \o{q}^{a f_1}_{\mu_1}({\bf x})
  \o{q}^{b f_2}_{\mu_2}({\bf x})
  \o{q}^{c' f_3}_{\mu_3}({\bf x}+\hat{\imath}) U^{\dagger c'c}_i({\bf x}),
  \label{eq:SD_general}
\end{equation}
where the time argument is omitted from quark fields, and
$\hat{\imath}$ is one of the six spatial directions $\{
{\pm\hat{\bf x},\pm\hat{\bf y},\pm\hat{\bf z}} \}$. Each quark
field is smeared but the third quark has an extra displacement by
one site from the other two in Eq.~(\ref{eq:SD_general}).

 Spatial displacements of Eq.~(\ref{eq:SD_general}) with
$\hat{\imath} \in \{ {\pm\hat{\bf x},\pm\hat{\bf y},\pm\hat{\bf
z}} \}$ transform amongst themselves under the rotations of the
octahedral group $O$ assuming that gauge links are cubically
invariant (approximately true for averages over large sets of
gauge-field configurations). The six-dimensional representation
of $O$ that is formed by the six displacements can be reduced to
the IRs $A_1$, $T_1$ and $E$. In order to combine displacements
so that they transform in the same way as the basis vectors of
$A_1, E, T_1$ IRs given in Table~\ref{table:lattice_harmonics},
we first define the following even and odd combinations of
forward and backward displacements:
 \begin{eqnarray}
\hat{S}_i\o{b}^{f_1 f_2 f_3}_{\mu_1 \mu_2 \mu_3} =
\o{b}^{( i) f_1 f_2 f_3}_{\mu_1 \mu_2 \mu_3} +
\o{b}^{(-i) f_1 f_2 f_3}_{\mu_1 \mu_2 \mu_3},
\nn \\
\hat{D}_i\o{b}^{f_1 f_2 f_3}_{\mu_1 \mu_2 \mu_3} =
\o{b}^{( i) f_1 f_2 f_3}_{\mu_1 \mu_2 \mu_3} -
\o{b}^{(-i) f_1 f_2 f_3}_{\mu_1 \mu_2 \mu_3},
\end{eqnarray}
with $i=x,y,z$. The difference of forward and backward
displacements, $\hat{D}_i$, has negative smearing-parity and
involves a lattice first-derivative, while the sum of forward and
backward displacements, $\hat{S}_i$, has positive parity. Note
that the lattice first-derivative is an anti-hermitian operator.
The second step is to form IR operators using the $\hat{S}_i$ and
$\hat{D}_i$ combinations as follows:
\begin{eqnarray}
\hat  A_1 \o{b}^{f_1 f_2 f_3}_{\mu_1 \mu_2 \mu_3} ({\bf x}) &\equiv&
  {1\over\sqrt{6}}\sum_{i=1,2,3} \hat{S}_i
  \o{b}^{f_1 f_2 f_3}_{\mu_1 \mu_2 \mu_3} ({\bf x}),
  \label{eq:one-link1} \\
\hat  E^1 \o{b}^{f_1 f_2 f_3}_{\mu_1 \mu_2 \mu_3} ({\bf x}) &\equiv&
  {1\over 2\sqrt{3}}\l(2\hat{S}_3 \o{b}^{f_1f_2f_3}_{\mu_1\mu_2\mu_3}({\bf x})-
  \hat{S}_1 \o{b}^{f_1 f_2 f_3}_{\mu_1 \mu_2 \mu_3} ({\bf x}) \r. \nn \\*
   & & \l. -\hat{S}_2 \o{b}^{f_1 f_2 f_3}_{\mu_1 \mu_2 \mu_3} ({\bf x}) \r),
  \label{eq:one-link2} \\
\hat  E^2 \o{b}^{f_1 f_2 f_3}_{\mu_1 \mu_2 \mu_3} ({\bf x}) &\equiv&
  {1\over 2} \l(
  \hat{S}_1 \o{b}^{f_1 f_2 f_3}_{\mu_1 \mu_2 \mu_3} ({\bf x}) -
  \hat{S}_2 \o{b}^{f_1 f_2 f_3}_{\mu_1 \mu_2 \mu_3} ({\bf x}) \r),
  \label{eq:one-link3} \\
\hat  T_1^1 \o{b}^{f_1 f_2 f_3}_{\mu_1 \mu_2 \mu_3} ({\bf x}) &\equiv&
  {i\over 2} \l(
  \hat{D}_x \o{b}^{f_1 f_2 f_3}_{\mu_1 \mu_2 \mu_3} ({\bf x}) + i
  \hat{D}_y \o{b}^{f_1 f_2 f_3}_{\mu_1 \mu_2 \mu_3} ({\bf x}) \r)
  \nonumber \\*
  &\equiv& \hat{D}^+\o{b}^{f_1 f_2 f_3}_{\mu_1 \mu_2 \mu_3} ({\bf x}),
  \label{eq:one-link4} \\
\hat  T_1^2 \o{b}^{f_1 f_2 f_3}_{\mu_1 \mu_2 \mu_3} ({\bf x}) &\equiv&
  -{i\over \sqrt{2}}\hat{D}_z \o{b}^{f_1 f_2 f_3}_{\mu_1 \mu_2 \mu_3}({\bf x})
  \nonumber \\*
  &\equiv& \hat{D}^0\o{b}^{f_1 f_2 f_3}_{\mu_1 \mu_2 \mu_3} ({\bf x}),
  \label{eq:one-link5} \\
\hat  T_1^3 \o{b}^{f_1 f_2 f_3}_{\mu_1 \mu_2 \mu_3} ({\bf x}) &\equiv&
  -{i\over 2} \l(
  \hat{D}_x \o{b}^{f_1 f_2 f_3}_{\mu_1 \mu_2 \mu_3} ({\bf x}) - i
  \hat{D}_y \o{b}^{f_1 f_2 f_3}_{\mu_1 \mu_2 \mu_3} ({\bf x}) \r)
  \nonumber \\*
  &\equiv& \hat{D}^-\o{b}^{f_1 f_2 f_3}_{\mu_1 \mu_2 \mu_3} ({\bf x}),
  \label{eq:one-link6}
\end{eqnarray}
These definitions produce spatial distributions
$\hat{A}_1,\hat{E}^\lambda,\hat{T}^\lambda_1$ that transform in
the same way as the
lattice harmonics of Table~\ref{table:lattice_harmonics}.
Superscripts on $\hat E$ and $\hat T_1$ operators refer to the
rows of the IRs. For the $\hat{T}_1^{1,2,3}$ combinations of
displacements, we will generally denote operators by using the
spherical notation $\hat{D}^{+,0,-}$ as defined by
Eqs.~(\ref{eq:one-link4}-\ref{eq:one-link6}).

We refer to these simplest nonlocal operators, involving linear
combinations of operators with the third quark field displaced by
one lattice site, as {\it one-link} operators. Let us denote the
general form of a one-link operator as
\begin{equation}
{\cal D}^{\Lambda}_{\lambda} \o{b}^{f_1 f_2 f_3}_{\mu_1 \mu_2 \mu_3} ({\bf x})
\equiv \eps_{a b c} \:
\o q^{a f_1}_{\mu_1}({\bf x})\o q^{b f_2}_{\mu_2}({\bf x})
\l[ {\cal D}_\lambda^\Lambda
\o q^{f_3}_{\mu_3}({\bf x}) \r]^c,
\label{eq:OneLink}
\end{equation}
where $\Lambda$ specifies the type of spatial IR ($A_1$, $T_1$
or $E$) and $\lambda$ specifies the row of the IR. In order to
combine the spatial IRs of the displacement operators with the
IRs of Dirac indices, we need the direct product rules.

\subsection{Direct products and Clebsch-Gordan coefficients}
\label{sec::Clebsch}

Nonlocal operators involve direct products of two different IRs
of the octahedral group, one associated with the combinations of
displacement operators and the other associated with the Dirac
indices. Linear combinations of such direct products can be
formed so that they transform irreducibly amongst themselves by
using Clebsch-Gordan coefficients for the octahedral group. These
have been published by Altmann and Herzig~\cite{Altmann}.

Clebsch-Gordan coefficients depend upon the basis of IR operators
but different choices of the bases are related to one another by
unitary transformations.  Because our basis operators differ from
those published by Altmann and Herzig, we have performed the
required unitary transformations and obtained suitable
Clebsch-Gordan coefficients for all possible direct products of
IRs of the double octahedral group.
A complete set of Clebsch-Gordan coefficients is given in
Appendix~\ref{app:Altmann} of this preprint and in
Ref.~\cite{IkuroThesis}.  The relative phases of operators from
different rows within an IR should be fixed in lattice
calculations in order to allow averaging over rows when that is
appropriate, as it is in mass calculations. However, different
ways of forming a given IR as direct products need not have the
same {\it overall} phases. We have used this freedom to eliminate
phases within each table of Clebsch-Gordan coefficients such that
all of our coefficients are real.

A one-link operator that transforms as overall IR $\Lambda$ and
row $\lambda$ of $O^D$ is written as a linear combination of
displacement operators acting on IRs of Dirac indices as
follows,
\begin{equation}
\o{\cal O}^{\Lambda}_\lambda = \sum_{\lambda_1,\lambda_2}
C \hspace{-1mm} \l(
{\footnotesize
\begin{array}{ccc}
\Lambda & \Lambda_1 & \Lambda_2 \\
\lambda & \lambda_1 & \lambda_2
\end{array} }
\hspace{-1mm} \r) {\cal D}^{\Lambda_1}_{\lambda_1} \o
\Psi^{\Lambda_2,k}_{\lambda_2}. \label{eq:CG_coefficient}
\end{equation}
where the corresponding quasi-local baryon operator is written as
$\o \Psi^{\Lambda,k}_{\lambda}$ instead of $\o
\Psi^{\Lambda,k}_{S,S_z}$ and the relation of $\lambda$ and
$S,S_z$ is obvious from Table~\ref{tab:spin_harmonics}. For
one-link operators, we need direct products of the IRs of
displacements ($\Lambda_1 = A_1, E$ and $T_1$) with the IRs of
Dirac indices of quasi-local baryon operators ($\Lambda_2 = G_1$
and $H$) . The following rules of group multiplication show which
overall IRs $\Lambda$ can be produced,
\begin{eqnarray}
A_1   \ot G_1   &=& G_1,     \nn \\
A_1   \ot \,H\, &=& H,       \nn \\
E     \ot G_1   &=& H,       \nn \\
E     \ot \,H\, &=& G_1 \oplus G_2 \oplus H,\nn \\
T_1   \ot G_1   &=& G_1 \oplus H,           \nn \\
T_1   \ot \,H\, &=& G_1 \oplus G_2 \oplus H \oplus H.
\label{eq:A1ET1GroupMultiplication}
\end{eqnarray}

\subsection{One-link operators}
\label{sec:One-Link}

Baryon operators with one-link displacements can be categorized
into two sets, one with antisymmetric and the other with
symmetric Dirac indices of the first two quarks. The
antisymmetric category includes the nucleon with MA isospin and
the $\Lambda$ baryon operators.  The symmetric category includes
the nucleon with MS isospin, and the $\Delta$, $\Omega$, $\Sigma$
and $\Xi$ baryon operators. These symmetries determine the
spinorial structures of the one-link operators.

One-link operators for the nucleon with MA isospin and for the
$\Lambda$ baryon are taken to be of the form,
\begin{eqnarray}
\hspace{-4mm} {\cal D}^{(3)\Lambda}_{\lambda}\o N^{\rm (MA)}_{\mu_1 \mu_2
\mu_3}  &=&  {\eps_{abc} \over \sqrt{2}}\l[
\o u_{\mu_1}^{a}  \o d_{\mu_2}^{b}  -
\o d_{\mu_1}^{a}  \o u_{\mu_2}^{b}  \r]
\l( {\cal D}^{\Lambda}_{\lambda} \o u_{\mu_3}  \r)^c  \hspace{-2mm},
\label{eq:uudMA_one-link}
\\
\hspace{-4mm}  {\cal D}^{(3)\Lambda}_{\lambda} \L_{\mu_1 \mu_2 \mu_3}  &=&
{\eps_{abc}\over \sqrt{2}}\l[
\o u_{\mu_1}^{a}  \o d_{\mu_2}^{b}  -
\o d_{\mu_1}^{a}  \o u_{\mu_2}^{b}  \r]
\l( {\cal D}^{\Lambda}_{\lambda} \o s_{\mu_3}  \r)^c  \hspace{-2mm},
\label{eq:Lambda_one-link}
\end{eqnarray}
where the superscript 3 of ${\cal D}^{(3)\Lambda}_{\lambda}$
denotes that the displacement operator ${\cal
D}^{\Lambda}_{\lambda} \in \{ \hat{A}_1, \hat{E}^\lambda,
\hat{T}_1^\lambda \}$ acts on the third quark, and $N^{\rm (MA)}$
denotes the isospin symmetry. This choice of one-link operators
preserves the antisymmetry under $\mu_1 \leftrightarrow \mu_2$,
and therefore requires Dirac indices to be MA (20 combinations)
or A (4 combinations). Taking into account the six possible
${\cal D}_{\lambda}^{\Lambda}$ combinations of displacements, the
total number of operators of the form of
Eq.~(\ref{eq:uudMA_one-link}) or Eq.~(\ref{eq:Lambda_one-link})
is $(20+4)\times 6=144$.

One-link operators for the nucleon with MS isospin, or for the
$\Delta$, $\Omega$, $\Sigma$ and $\Xi$ baryons have the following
forms:
\begin{eqnarray}
{\cal D}_{\lambda}^{\Lambda (3)} \o N^{\rm (MS)}_{\mu_1 \mu_2 \mu_3} &=&
{\eps_{abc}\over \sqrt{6}} \l[ 2
\o u_{\mu_1}^a \o u_{\mu_2}^b
\l( {\cal D}^{\Lambda}_{\lambda}\o d_{\mu_3}\r)^c \r. \nn \\*
&-& \l. \l( \o u_{\mu_1}^a \o d_{\mu_2}^b + \o d_{\mu_1}^a \o u_{\mu_2}^b \r)
\l( {\cal D}^{\Lambda}_{\lambda} \o u_{\mu_3}\r)^c \r],
\label{eq:uudMS_one-link}
\\
{\cal D}_{\lambda}^{\Lambda (3)} \o \Delta^{++}_{\mu_1 \mu_2 \mu_3} &=&
\eps_{abc}\: \o u_{\mu_1}^a \o u_{\mu_2}^b
\l( {\cal D}^{\Lambda}_{\lambda}\o u_{\mu_3}\r)^c,
\label{eq:Delta_one-link}
\\
{\cal D}_{\lambda}^{\Lambda (3)} \o \Omega^-_{\mu_1 \mu_2 \mu_3} &=&
\eps_{abc}\: \o s_{\mu_1}^a \o s_{\mu_2}^b
\l( {\cal D}^{\Lambda}_{\lambda}\o s_{\mu_3}\r)^c,
\label{eq:Omega_one-link}
\\
{\cal D}_{\lambda}^{\Lambda (3)} \o \Sigma^+_{\mu_1 \mu_2 \mu_3} &=&
\eps_{abc}\: \o u_{\mu_1}^a \o u_{\mu_2}^b
\l( {\cal D}^{\Lambda}_{\lambda}\o s_{\mu_3}\r)^c,
\label{eq:Sigma_one-link}
\\
{\cal D}_{\lambda}^{\Lambda (3)} \o \Xi^0_{\mu_1 \mu_2 \mu_3} &=&
\eps_{abc}\: \o s_{\mu_1}^a \o s_{\mu_2}^b
\l( {\cal D}^{\Lambda}_{\lambda}\o u_{\mu_3}\r)^c.
\label{eq:Cascade_one-link}
\end{eqnarray}
These operators are symmetric under $\mu_1 \leftrightarrow \mu_2$, so
the allowed combinations of Dirac indices are
totally symmetric (20 combinations)
or mixed-symmetric (20 combinations). There are $(20+20) \times 6 = 240$
such operators for each baryon.

\subsubsection{$A_1$ one-link operators}
\label{sec:A1One-Link}

The reduction is the simplest for the $\hat A_1$ combination of
one-link operators because it is just a scalar ``smearing''.   We
show it as a first example. The MA isospin nucleon operator of
Eq.~(\ref{eq:uudMA_one-link}) and the $\Lambda$ baryon operator of
Eq.~(\ref{eq:Lambda_one-link}) have the same restriction on Dirac
indices as in Eq.~(\ref{eq:restriction_lambda}). Because the
$A_1$ combination of displacements is cubically symmetric, these
operators have the same transformations under group rotations as
the quasi-local $\Lambda$ baryon operators in
Eq.~(\ref{eq:LocalLambda}), except that the strange quark is
replaced by $(\hat{A}_1 \o u_{\mu_3})^c$ and $(\hat{A}_1 \o
s_{\mu_3})^c$, respectively. Dirac indices for $\hat{A}^{(3)}_1
\o N^{\rm (MA)}$ and $\hat{A}^{(3)}_1 \L$ are obtained from
Table~\ref{table:LocalLambda}, the quasi-local $\Lambda$ baryon
operator table.

For each operator in
Eqs.~(\ref{eq:uudMS_one-link})-(\ref{eq:Cascade_one-link}), the
displacement makes the third quark distinct but the operators are
symmetric under $\mu_1 \leftrightarrow \mu_2$ as in
Eq.~(\ref{eq:restriction_sigma}). This means that these operators
transform in the same manner as the quasi-local $\Sigma$ baryon
operators and Table~\ref{table:LocalSigma} can be used for any of
the operators in
Eqs.~(\ref{eq:uudMS_one-link})-(\ref{eq:Cascade_one-link}).

We note in passing that any cubically symmetric form of smearing
can be developed by repeated application of the $\hat A_1$
operator. Thus, any such smearing that makes the spatial
distribution of the third quark different from that of the first
two can be substituted for the $\hat{A}_1$ combination of
displacements of the third quark. All such operators have the
same transformations and thus the same IRs of Dirac indices.

\begin{table*}
\caption{$T_1$ one-link operators. Note that
         $\hat{D}^+ \equiv \hat{T}_1^1$,
         $\hat{D}^0 \equiv \hat{T}_1^2$, and
         $\hat{D}^- \equiv \hat{T}_1^3$.}
\label{table:One-LinkPWave}
\begin{center}
\renewcommand{\arraystretch}{1.4} 
\begin{tabular}{|cccc|}
\hline
IR & row &
$C \hspace{-1mm} \l(
{\footnotesize
\renewcommand{\arraystretch}{0.01}
\begin{array}{ccc}
\vspace{-1mm}
\Lambda &    T_1     &  \Lambda_2 \\
\lambda & \lambda_1  &  \lambda_2
\end{array} }
\hspace{-1mm} \r) \hat{D}^{(3) \lambda_1} \o
\Psi^{\Lambda_2,k}_{\lambda_2}$
      &  $\sim \l| J, J_z \r>$ \\
\hline \hline $H_{g/u}$ & 1 & $\hat{D}^+ \o \Psi^{G_{1u/g},
k}_{{1\over2},{1\over2}}$   &
  $\l| {3\over2},+{3\over2} \r>$ \\
  &2& ${1\over \sqrt{3}} \hat{D}^+ \o \Psi^{G_{1u/g}, k}_{{1\over2},-{1\over2}}
  +\sqrt{2\over 3}\hat{D}^0 \o \Psi^{G_{1u/g}, k}_{{1\over2},{1\over2}}$   &
  $\l| {3\over2},+{1\over2} \r>$ \\
  & 3 & $\sqrt{2\over 3}\hat{D}^0 \o \Psi^{G_{1u/g}, k}_{{1\over2},-{1\over2}}
  + {1\over \sqrt{3}} \hat{D}^- \o \Psi^{G_{1u/g}, k}_{{1\over2},{1\over2}}$  &
  $\l| {3\over2},-{1\over2} \r>$ \\
  & 4 & $\hat{D}^- \o \Psi^{G_{1u/g}, k}_{{1\over2},-{1\over2}}$   &
  $\l| {3\over2},-{3\over2} \r>$ \\
\hline
$H_{g/u}$ & 1 &
 $ \sqrt{3\over 5}\hat{D}^0 \o \Psi^{H_{u/g}, k}_{{3\over2},{3\over2}}
 -{\sqrt{2 \over 5}} \hat{D}^+ \o \Psi^{H_{u/g}, k}_{{3\over2},{1\over2}}$  &
 $\l| {3\over2},+{3\over2} \r>$ \\
 & 2 & $ \sqrt{2\over 5}\hat{D}^- \o \Psi^{H_{u/g}, k}_{{3\over2},{3\over2}}
 + {1 \over \sqrt{15}} \hat{D}^0 \o \Psi^{H_{u/g}, k}_{{3\over2},{1\over2}}
 - {\sqrt{8 \over 15}} \hat{D}^+ \o \Psi^{H_{u/g}, k}_{{3\over2},-{1\over2}}$&
 $\l| {3\over2},+{1\over2} \r>$ \\
 & 3 & $ \sqrt{8\over 15}\hat{D}^- \o \Psi^{H_{u/g}, k}_{{3\over2},{1\over2}}
 - {1 \over \sqrt{15}} \hat{D}^0 \o \Psi^{H_{u/g}, k}_{{3\over2},-{1\over2}}
 - {\sqrt{2 \over 5}} \hat{D}^+ \o \Psi^{H_{u/g}, k}_{{3\over2},-{3\over2}}$  &
 $\l| {3\over2},-{1\over2} \r>$ \\
 & 4 &${\sqrt{2 \over 5}} \hat{D}^- \o \Psi^{H_{u/g}, k}_{{3\over2},-{1\over2}}
 -  \sqrt{3\over 5}\hat{D}^0 \o \Psi^{H_{u/g}, k}_{{3\over2},-{3\over2}}$   &
 $\l| {3\over2},-{3\over2} \r>$ \\
\hline
$H_{g/u}$ & 1 &
 ${1\over \sqrt{10}} \hat{D}^+ \o \Psi^{H_{u/g},k}_{{3\over2},{1\over2}}
 +  \sqrt{5\over 6}\hat{D}^- \o \Psi^{H_{u/g}, k}_{{3\over2},-{3\over2}}
 + {1 \over \sqrt{15}} \hat{D}^0 \o \Psi^{H_{u/g}, k}_{{3\over2},{3\over2}}$&
 ${1\over\sqrt{6}}\l| {5\over2},+{3\over2} \r>
 +\sqrt{5\over6}  \l| {5\over2},-{5\over2} \r>$ \\
 & 2 &$ -{1\over \sqrt{10}}\hat{D}^- \o \Psi^{H_{u/g}, k}_{{3\over2},{3\over2}}
 - \sqrt{3\over 5} \hat{D}^0 \o \Psi^{H_{u/g}, k}_{{3\over2},{1\over2}}
 - {\sqrt{3 \over 10}} \hat{D}^+ \o \Psi^{H_{u/g}, k}_{{3\over2},-{1\over2}}$&
 $\l| {5\over2},+{1\over2} \r>$  \\
 & 3 & $ \sqrt{3\over 10}\hat{D}^- \o \Psi^{H_{u/g}, k}_{{3\over2},{1\over2}}
 + \sqrt{3\over 5} \hat{D}^0 \o \Psi^{H_{u/g}, k}_{{3\over2},-{1\over2}}
 + {1\over \sqrt{10}} \hat{D}^+ \o \Psi^{H_{u/g}, k}_{{3\over2},-{3\over2}}$&
 $\l| {5\over2},-{1\over2} \r>$  \\
 & 4 &$-{\sqrt{5 \over 6}} \hat{D}^+ \o \Psi^{H_{u/g}, k}_{{3\over2},{3\over2}}
 - {1\over \sqrt{10}}\hat{D}^- \o \Psi^{H_{u/g}, k}_{{3\over2},-{1\over2}}
 - {1\over \sqrt{15}} \hat{D}^0\o \Psi^{H_{u/g}, k}_{{3\over2},-{3\over2}}$&
 $\sqrt{5\over6}   \l| {5\over2},+{5\over2} \r>
 +{1\over\sqrt{6}} \l| {5\over2},-{3\over2} \r>$ \\
\hline
$G_{1g/u}$ & 1 &
 $\sqrt{2\over 3} \hat{D}^+\o \Psi^{G_{1u/g},k}_{{1\over2},-{1\over2}}
 - {1\over \sqrt{3}}\hat{D}^0 \o \Psi^{G_{1u/g}, k}_{{1\over2},{1\over2}}$   &
 $\l| {1\over2},+{1\over2} \r>$ \\
 & 2 & $ {1\over \sqrt{3}}\hat{D}^0 \o \Psi^{G_{1u/g},k}_{{1\over2},-{1\over2}}
 - \sqrt{2\over 3} \hat{D}^- \o \Psi^{G_{1u/g}, k}_{{1\over2},{1\over2}}$   &
 $\l| {1\over2},-{1\over2} \r>$ \\
\hline
$G_{1g/u}$ & 1 &
 ${1\over \sqrt{2}} \hat{D}^- \o \Psi^{H_{u/g}, k}_{{3\over2},{3\over2}}
 -  {1\over \sqrt{3}}\hat{D}^0 \o \Psi^{H_{u/g}, k}_{{3\over2},{1\over2}}
 +  {1\over \sqrt{6}}\hat{D}^+ \o \Psi^{H_{u/g}, k}_{{3\over2},-{1\over2}}$  &
 $\l| {1\over2},+{1\over2} \r>$ \\
 & 2 & $ {1\over \sqrt{6}}\hat{D}^- \o \Psi^{H_{u/g}, k}_{{3\over2},{1\over2}}
 -  {1\over \sqrt{3}}\hat{D}^0 \o \Psi^{H_{u/g}, k}_{{3\over2},-{1\over2}}
 + {1\over \sqrt{2}} \hat{D}^+ \o \Psi^{H_{u/g}, k}_{{3\over2},-{3\over2}}$  &
 $\l| {1\over2}, -{1\over2} \r>$ \\
\hline
$G_{2g/u}$ & 1 &
 $ -{1\over \sqrt{2}}\hat{D}^- \o \Psi^{H_{u/g}, k}_{{3\over2},-{1\over2}}
 -  {1\over \sqrt{3}}\hat{D}^0 \o \Psi^{H_{u/g}, k}_{{3\over2},-{3\over2}}
 + {1\over \sqrt{6}} \hat{D}^+ \o \Psi^{H_{u/g}, k}_{{3\over2},{3\over2}}$  &
 ${1\over\sqrt{6}} \l| {5\over2},+{5\over2} \r>
 - \sqrt{5\over6}   \l| {5\over2},-{3\over2} \r>$ \\
 & 2 &
 ${1\over \sqrt{6}} \hat{D}^- \o \Psi^{H_{u/g}, k}_{{3\over2},-{3\over2}}
 -  {1\over \sqrt{3}}\hat{D}^0 \o \Psi^{H_{u/g}, k}_{{3\over2},{3\over2}}
 -  {1\over \sqrt{2}}\hat{D}^+ \o \Psi^{H_{u/g}, k}_{{3\over2},{1\over2}}$  &
 $-\sqrt{5\over6}  \l| {5\over2},+{3\over2} \r>
 +{1\over\sqrt{6}} \l| {5\over2},-{5\over2} \r>$ \\
\hline
\end{tabular}
\end{center}
\end{table*}
\begin{table}
\begin{center}
\caption{$E$ one-link operators. All operators have mixed $J_z$.}
\label{table:One-LinkDWave}
\renewcommand{\arraystretch}{1.5} 
\begin{tabular}{|ccc|}
\hline
IR & row &
$C \hspace{-1mm} \l(
{\footnotesize
\renewcommand{\arraystretch}{0.01}
\begin{array}{ccc}
\vspace{-1mm}
\Lambda &     E      &  \Lambda_2 \\
\lambda & \lambda_1  &  \lambda_2
\end{array} }
\hspace{-1mm}
\r)
\hat{E}^{(3) \lambda_1} \o \Psi^{\Lambda_2,k}_{\lambda_2}$ \\
\hline
\hline
$H_{g/u}$ & 1 & $-\hat E^2\o \Psi^{G_{1g/u},k}_{{1\over 2},-{1\over 2}}$ \\
          & 2 & $ \hat E^1\o \Psi^{G_{1g/u},k}_{{1\over 2},{1\over 2}}$  \\
          & 3 & $-\hat E^1\o \Psi^{G_{1g/u},k}_{{1\over 2},-{1\over 2}}$ \\
          & 4 & $ \hat E^2\o \Psi^{G_{1g/u},k}_{{1\over 2},{1\over 2}}$  \\
\hline
$H_{g/u}$ & 1 & $-\hat E^1\o \Psi^{H_{g/u},k}_{{3\over 2},{3\over 2}}
          - \hat E^2\o \Psi^{H_{g/u},k}_{{3\over 2},-{1\over 2}}$ \\
      & 2 & $\hat E^1\o \Psi^{H_{g/u},k}_{{3\over 2},{1\over 2}}
      - \hat E^2\o \Psi^{H_{g/u},k}_{{3\over 2},-{3\over 2}}$ \\
      & 3 & $\hat E^1\o \Psi^{H_{g/u},k}_{{3\over 2},-{1\over 2}}
      - \hat E^2\o \Psi^{H_{g/u},k}_{{3\over 2},{3\over 2}}$ \\
      & 4 & $-\hat E^1\o \Psi^{H_{g/u},k}_{{3\over 2},-{3\over 2}}
      - \hat E^2\o \Psi^{H_{g/u},k}_{{3\over 2},{1\over 2}}$ \\
\hline
$G_{1g/u}$& 1 & $\hat E^1\o \Psi^{H_{g/u},k}_{{3\over 2},{1\over 2}}
          + \hat E^2\o \Psi^{H_{g/u},k}_{{3\over 2},-{3\over 2}}$ \\
          & 2 & $-\hat E^1\o \Psi^{H_{g/u},k}_{{3\over 2},-{1\over 2}}
          - \hat E^2\o \Psi^{H_{g/u},k}_{{3\over 2},{3\over 2}}$ \\
\hline
$G_{2g/u}$& 1 & $\hat E^1\o \Psi^{H_{g/u},k}_{{3\over 2},-{3\over 2}}
          - \hat E^2\o \Psi^{H_{g/u},k}_{{3\over 2},{1\over 2}}$  \\
      & 2 & $-\hat E^1\o \Psi^{H_{g/u},k}_{{3\over 2},{3\over 2}}
      + \hat E^2\o \Psi^{H_{g/u},k}_{{3\over 2},-{1\over 2}}$ \\
\hline
\end{tabular}
\end{center}
\end{table}
%

\subsubsection{$T_1, E$ one-link operators}
\label{sec:T1E_One-Link}

  In order to construct operators that have the $T_1$
or $E$ combinations of one-link displacements, we apply the
Clebsch-Gordan formula of Eq.~(\ref{eq:CG_coefficient}) using the
coefficients for the double octahedral group from
Appendix~\ref{app:Altmann}.  The resulting one-link operators are
given in Table~\ref{table:One-LinkPWave} and
Table~\ref{table:One-LinkDWave}.  These tables give {\it all}
possible $T_1$ and $E$ one-link baryon operators. The parity
labels refer to the intermediate states that the operator couples
with in a zero-momentum correlation function.  In
Table~\ref{table:One-LinkPWave} we employ the notation of
$\hat{D}^{+,-,0}$, instead of $\hat{T}_1^{1,2,3}$. The
displacements are understood to act on the third quark.  These
tables are general in the sense that they apply to any baryon,
e.g., $\o N,\D ,\L, \S$, or $\o \Xi$.

The notation $\o\Psi^{\Lambda, k}_\lambda$ describes a
\textit{quasi-local} operator whose spin belongs to the $k$-th
embedding of IR $\Lambda$ and row $\lambda$.  These operators
are taken directly from the tables for quasi-local baryon
operators discussed in Section~\ref{sec:local} in a similar
fashion as for the $A_1$ one-link operators. One-link nucleon
operators with MA isospin and one-link $\Lambda$ baryon operators
employ the spinorial structures $\o \Psi^{\Lambda, k}_\lambda$
used for quasi-local $\Lambda$ baryon operators given in
Table~\ref{table:LocalLambda}, together with
Table~\ref{table:One-LinkPWave} for $T_1$ and
Table~\ref{table:One-LinkDWave} for $E$. One-link nucleon
operators with MS isospin or one-link $\Delta$, $\Omega$ $\Sigma$
and $\Xi$ operators employ the same spinorial structures $\o
\Psi^{\Lambda, k}_\lambda$ that are given for the quasi-local
$\Sigma$ baryon operators in Table~\ref{table:LocalSigma}.

The $T_1$ one-link operators in
Table~\ref{table:One-LinkPWave} are strictly ``barred'' fields.
The corresponding ``unbarred'' $T_1$ operators use hermitian
conjugated lattice first-derivatives.
The factor $i$ that has been included in the $T_1$ one-link
operators provides the same hermiticity property as spherical
harmonics, i.e., $Y_{l,m}^\dag=(-1)^m Y_{l,-m}$. Note that
because the smearing parity of the $T_1$ displacement is negative,
the overall $\rho$-parity is opposite to the overall parity for
$T_1$ one-link operators.

The last column of Table~\ref{table:One-LinkPWave} shows the total
angular momentum and its projection onto the $z$-axis. The first
set of $H$ operators involves the same Clebsch-Gordan
coefficients as would apply to the formation of continuum states:
$(L=1) \ot (S=1/2) \rightarrow (J=3/2)$. This is a result of
using the combinations of displacements that transform in the
same way as the basis vectors of
Table~\ref{table:lattice_harmonics}. The second set of $H$
operators has the continuum Clebsch-Gordan coefficients for
$(L=1) \ot (S=3/2) \rightarrow (J=3/2)$. The third set of $H$
operators corresponds to $(L=1) \ot (S=3/2) \rightarrow (J=5/2)$,
but $J_z$ values are mixed in row 1 and row 4. Similarly, the
first set of $G_1$ operators in Table~\ref{table:One-LinkPWave}
is $(L=1) \ot (S=1/2) \rightarrow (J=1/2)$, and the second one is
$(L=1) \ot (S=3/2) \rightarrow (J=1/2)$, both having the
continuum Clebsch-Gordan coefficients. The $G_2$ operators
correspond to $(L=1) \ot (S=3/2) \rightarrow (J=5/2)$, but $J_z$
values are mixed.

Direct products involving the $\hat E$ spatial IR of
displacements and spinorial IRs are given in
Table~\ref{table:One-LinkDWave}. No operators involve {\it
continuum} Clebsch-Gordan coefficients in
Table~\ref{table:One-LinkDWave} because the $E$ IR has mixed
$J_z$, i.e., $\hat{E}^2 \sim Y_{2,2}+Y_{2-2}$. The $\hat{E}$
combinations of displacements provide two members of the
rank-two spherical harmonics. The remaining three members belong
to the $T_2$ IR and they cannot be constructed unless there
are least two displacements in perpendicular directions, as will
be discussed in the next section.

For baryon fields with projection to zero total momentum, the
following linear dependence holds,
\begin{eqnarray}
&&\sum_x
\o{q}_{\mu_1} ({\bf x}) \o{q}_{\mu_2} ({\bf x}) \hat{D}^i \o{q}_{\mu_3}({\bf x})
\nonumber \\
= -&&\sum_x \o{q}_{\mu_1} ({\bf x}) \l(\hat{D}^i \o{q}_{\mu_2}({\bf x})\r)
\o{q}_{\mu_3} ({\bf x}) \nn \\*
 - &&\sum_x \l(\hat{D}^i \o{q}_{\mu_1} ({\bf x})\r)
\o{q}_{\mu_2} ({\bf x})\o{q}_{\mu_3} ({\bf x}) + O(a^2).~
\label{eq:TotalDerivative}
\end{eqnarray}
This relation derives from the fact that after projection to
zero total momentum, a total derivative of a
baryon field vanishes and a total derivative is equivalent
to order $a^2$ to a sum of lattice derivatives applied to each
quark field. Some of the $T_1$ one-link nucleon operators are not
linearly independent because of this. A nucleon operator with MS
isospin having MS Dirac indices is equivalent (within a total
derivative) to a nucleon operator with MA isospin having MA Dirac
indices, for the $T_1$ one-link construction. It is easy to show
that $\hat{D}^{(3)}_i \o N^{\rm (MS)}_{\mu_1 \mu_2 \mu_3} +
{1\over 2} (\mu_2\leftrightarrow\mu_3) + {1\over 2}
(\mu_1\leftrightarrow\mu_3)$ can be written as a linear
combination of $\hat{D}^{(3)}_i \o N^{\rm
(MA)}_{\mu_1 \mu_2 \mu_3}$'s by applying
Eq.~(\ref{eq:TotalDerivative}).  This identity reduces the number
of $T_1$ one-link nucleon operators by $20\times 3 = 60$, where
the number 20 comes from the number of MS Dirac indices (or MA
Dirac indices). The number of distinct one-link nucleon operators
(both MA and MS isospin) after projection to zero total momentum
is 64 for $A_1$, 132 for $T_1$, and 128 for $E$.  The total
number is 324.

Operators that are totally symmetric with respect to flavor
exchanges, such as the $\Delta$ baryon, have a similar
restriction. Such operators vanish when a first-derivative acts
on one quark in a totally symmetric combination of Dirac indices.
There are sixty $\Delta$ baryon operators with the $T_1$ one-link
displacements that vanish after projection to zero total momentum.

\begin{table}[h]
\caption{Allowed combinations of Dirac indices for different
one-link ($A_1,T_1,E$) baryons.  The displacement is always taken
on the third quark for simplicity.  The third quark of the
$\Lambda$ and $\Sigma$ baryons is chosen to be strange quark, and
the third quark of the $\Xi$ baryon is the light quark. The
numbers of operators for $A_1$, $T_1$, or $E$ combinations of
displacements are listed in the fifth column and the numbers of
operators for each overall IR of $O^D$ are shown in the last
three columns, counting both parities, all embeddings and all
rows. Linear dependencies resulting from a projection to zero
total momentum are not taken into account in this table.}
\begin{center}
\begin{ruledtabular}
\begin{tabular}{ccccc|ccc}
one-link baryon & spin sym. & Table &
disp. & \# & $G_1$ & $G_2$ & $H$ \\
\hline \hline $N^{\rm (MA)},\Lambda$ & MA, A
&\ref{table:LocalLambda}                             & $A_1$ &24&16&0& 8 \\
& &\ref{table:One-LinkPWave}, \ref{table:LocalLambda}& $T_1$ &72&20&4&48 \\
& &\ref{table:One-LinkDWave}, \ref{table:LocalLambda}&  $E$  &48& 4&4&40 \\
\hline
$N^{\rm (MS)},\Delta,\Omega,\Sigma,\Xi$ & MS, S &\ref{table:LocalSigma}
                                                    & $A_1$ & 40&16& 0&24 \\
& &\ref{table:One-LinkPWave},\ref{table:LocalSigma} & $T_1$ &120&28&12&80 \\
& &\ref{table:One-LinkPWave},\ref{table:LocalSigma} &  $E$  & 80&12&12&56 \\
\end{tabular}
\end{ruledtabular}
\end{center}
\label{table:One-LinkIsospinDiracSymmetry}
\end{table}
The correspondence between the type of baryon and the symmetry of
Dirac indices for the two categories of one-link baryon operators
is summarized in Table~\ref{table:One-LinkIsospinDiracSymmetry}.
The numbers of possible operators are shown for constructions
using $A_1$, $T_1$, or $E$ spatial IRs to obtain $G_1$, $G_2$,
or $H$ overall IRs.

\subsection{Two-link operators}
\label{sec:TwoLink}

One-link operators make it possible to realize $\hat A_1$, $\hat
T_1$ and $\hat E$ types of spatial smearing, but not the $\hat
T_2$ or $\hat A_2$ types. The latter two types appear in the
two-link operator constructions. We define a {\it two-link
operator} as follows,
\begin{eqnarray}
&& \hspace{-8mm}
{\cal D}^{(3) \Lambda_2}_{\lambda_2}{\cal D}^{(3) \Lambda_1}_{\lambda_1}
\o b^{f_1f_2f_3}_{\mu_1\mu_2\mu_3} ({\bf x}) \equiv \nn \\*
&& \eps_{a b c}
\o q^{a f_1}_{\mu_1}({\bf x})\o q^{b f_2}_{\mu_2}({\bf x})
\l[ {\cal D}_{\lambda_2}^{\Lambda_2}{\cal D}_{\lambda_1}^{\Lambda_1}
\o q^{f_3}_{\mu_3}({\bf x}) \r]^c,
\label{eq:TwoLink}
\end{eqnarray}
where the third quark is displaced covariantly by two
displacement operators ${\cal D}_{\lambda_1}^{\Lambda_1},{\cal
D}^{\Lambda_2}_{\lambda_2} \in \{A_1,E^\lambda,T_1^\lambda\}$.
The first displacement acts on the third quark and defines a
modified quark field, ${\tilde {q}}^{c' f_3, \Lambda_1}_{\mu_3,
\lambda_1}({\bf x}) \equiv \l[ {\cal D}^{\Lambda_1}_{\lambda_1}
\o{q}^{f_3}_{\mu_3}({\bf x})\r]^{c'}$ at the same position ${\bf
x}$. Then the second displacement further displaces the field and
so defines a second modified field at the same position,
${\tilde{\tilde{q}}}^{c
f_3,\Lambda_1,\Lambda_2}_{\mu_3,\lambda_1,\lambda_2} ({\bf x})
\equiv \l[ {\cal D}^{\Lambda_2}_{\lambda_2} {\tilde {q}}^{f_3,
\Lambda_1}_{\mu_3, \lambda_1}({\bf x})\r]^c$.

\begin{figure}[h]
\centering
\includegraphics[angle=0,width=50mm]{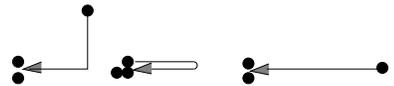}
\caption{Illustration of two-link displacements.}
\label{fig:TwoLinkFig}
\end{figure}

Figure~\ref{fig:TwoLinkFig} shows schematic illustrations of
three distinct displacement configurations for a two-link baryon
operator. The first figure shows the ``bent-link'' operator,
where a line denotes the gauge link and the arrow specifies the
point at which the displaced quark's color index forms a color
singlet with the other quarks. The second figure shows the
possibility that the third quark is translated back to its
original position by the second displacement, which is equivalent
to a quasi-local operator because $U_i(x)U^\dag_i(x)=1$. The
third figure shows the possibility of two displacements along the
same direction, which gives a straight path differing from a
one-link displacement only by its length.
Inclusion of the bent-links can enrich the angular distribution and recover
parts of the continuum spherical harmonics that cannot be obtained from
one-link displacements.

First, we classify the spatial degrees of freedom into a single
IR of $O$ by forming linear combinations of the elemental
operators of Eq.~(\ref{eq:TwoLink}). The overall spatial IR
$\Lambda$ and row $\lambda$ are determined by the direct product
of the two spatial displacements ${\cal
D}_{\lambda_1}^{\Lambda_1}$ and ${\cal
D}_{\lambda_2}^{\Lambda_2}$ with appropriate Clebsch-Gordan
coefficients,
\begin{equation}
\sum_{\lambda_1,\lambda_2}
C \hspace{-1mm} \l(
{\footnotesize
\begin{array}{ccc}
\Lambda & \Lambda_1 & \Lambda_2 \\
\lambda & \lambda_1 & \lambda_2
\end{array} }
\hspace{-1mm} \r) {\cal D}^{\Lambda_2}_{\lambda_2} {\cal
D}^{\Lambda_1}_{\lambda_1}.
 \label{eq:CG_coefficient_DD}
\end{equation}
A particular example is instructive.  Suppose one chooses
${\cal D}_{\lambda_1}^{\Lambda_1}$ to belong to the $T_1$ IR, and
${\cal D}_{\lambda_2}^{\Lambda_2}$ to belong to the $E$ IR
and desires the overall spatial IR to be $T_1$. Then
Eq.(\ref{eq:CG_coefficient_DD}) is used with Clebsch-Gordan
coefficients from the $E\ot T_1$ table in
Appendix~\ref{app:Altmann}, which gives
\begin{eqnarray}
\hat{\hat{T}}_1^1 &\sim&  {1\over2}\hat{E}^1 \times \hat{T}_1^1
                   + {\sqrt{3}\over2} \hat{E}^2 \times \hat{T}_1^3, \nn \\
\hat{\hat{T}}_1^2 &\sim& - \hat{E}^1 \times \hat{T}_1^2,\nn \\
\hat{\hat{T}}_1^3 &\sim&  {1\over2}\hat{E}^1 \times \hat{T}_1^3
                   + {\sqrt{3}\over2} \hat{E}^2 \times \hat{T}_1^1.\nn
\end{eqnarray}
In this way the two-link operator is determined so that its
spatial part transforms according to a particular IR (in this
case $T_1$). Once the overall spatial IR is obtained,
Clebsch-Gordan coefficients for the direct products of the
overall spatial IR and a selected spinorial IR are used to
form an operator that overall transforms irreducibly according to
$G_1,H$, or $G_2$. Because the spatial IR in the example above
is $T_1$, which has been considered already in the construction
of one-link operators, Table~\ref{table:One-LinkPWave} provides
the result. The only change is to use the $T_1$ distribution of
two-link displacements in place of the $T_1$ distribution of
one-link displacements. The use of Clebsch-Gordan coefficients of
the cubic group has reduced the problem of finding IRs of
two-link baryon operators to the already solved problem of finding
one-link baryon operators.

However, new possibilities exist with the two-link displacements.
One can form the $A_2$ and $T_2$ spatial IRs that did not appear
in the one-link construction. This construction is
straightforward but is omitted from this paper except to note that
two-link $A_2$ and $T_2$ lattice harmonics correspond to the
spherical harmonics shown in Table~\ref{table:lattice_harmonics}.

Proceeding in this fashion, one may construct {\it multi-link}
baryon operators
\begin{eqnarray}
&& {\cal D}^{(3)\Lambda_n }_{\lambda_n} \cdot \cdot \cdot {\cal
D}^{(3)\Lambda_2}_{\lambda_2} {\cal D}^{(3) \Lambda_1}_{\lambda_1}
\o b^{f_1 f_2 f_3}_{\mu_1\mu_2\mu_3} ({\bf x}) \equiv \nn \\*
&& \eps_{a b c}
\o q^{a f_1}_{\mu_1}({\bf x})\o q^{b f_2}_{\mu_2}({\bf x}) \l[ {\cal
D}_{\lambda_n}^{\Lambda_n} \cdot \cdot \cdot {\cal
D}_{\lambda_2}^{\Lambda_2}{\cal D}_{\lambda_1}^{\Lambda_1} \o
q^{f_3}_{\mu_3}({\bf x}) \r]^c,
\label{eq:ThreeLink}
\end{eqnarray}
that involve $n$-site displacements in space allowing a quark to
be displaced over a finer angular distribution so as to yield
higher rank spherical harmonics. The reduction procedure is
essentially the same as for the two-link case, except that
multiple direct products of spatial IRs are used.

\subsection{One-link displacements applied to two different quarks}
\label{sec:TwoQuarkDisplacement}

Consider an operator with one-link displacements applied to two
different quarks in the following way,
\begin{eqnarray}
&&\hspace{-10mm}
{\cal D}^{(1\pm2) \Lambda_2}_{\lambda_2}{\cal D}^{(3) \Lambda_1}_{\lambda_1}
\o b^{f_1f_2f_3}_{\mu_1\mu_2\mu_3}
\equiv \nn \\*
&&\hspace{-10mm} {\eps_{a b c}\over\sqrt{2}}
\l[ \l( {\cal D}_{\lambda_2}^{\Lambda_2} \o q^{f_1}_{\mu_1} \r)^a
\o q^{b f_2}_{\mu_2} \pm
\o q^{a f_1}_{\mu_1}
\l({\cal D}_{\lambda_2}^{\Lambda_2}\o q^{f_2}_{\mu_2} \r)^b \r]
\l({\cal D}_{\lambda_1}^{\Lambda_1}\o q^{f_3}_{\mu_3} \r)^c \hspace{-2mm},
\label{eq:TwoQuarkDisp}
\end{eqnarray}
where ${\cal D}^{(1\pm2) \Lambda}_\lambda$ indicates that the
first two quark fields are symmetric or antisymmetric
with respect to exchange of their spatial dependencies,
\begin{equation}
{\cal D}^{(1\pm 2) \Lambda}_\lambda =
{1\over\sqrt{2}} \l( {\cal D}^{(1)\Lambda}_\lambda
\pm {\cal D}^{(2)\Lambda}_\lambda \r).
\end{equation}
We refer to ${\cal D}^{(1+2)\Lambda }_\lambda$ as {\it
space-symmetric} and to ${\cal D}^{(1-2)\Lambda }_\lambda$ as
{\it space-antisymmetric} combinations of displacements.  The
symmetry of the spatial displacements must be taken into account
in the overall antisymmetry of operators in order to identify the
symmetry of Dirac indices that produces nonvanishing operators.

For the case of MA isospin nucleon operators with one-link displacements
applied to two quarks, we obtain
\begin{eqnarray}
&& \hspace{-9mm}
{\cal D}^{(1\pm2) \Lambda_2}_{\lambda_2}{\cal D}^{(3) \Lambda_1}_{\lambda_1}
\o N^{\rm (MA)}_{\mu_1\mu_2\mu_3} \nn \\*
&=& \eps_{a b c} {1\over 2}
\l( {\cal D}_{\lambda_2}^{\Lambda_2} \o{u}^a_{\mu_1}
                   \o{d}^b_{\mu_2} -
    {\cal D}_{\lambda_2}^{\Lambda_2} \o{d}^a_{\mu_1}
                   \o{u}^b_{\mu_2} \r)
 \l({\cal D}_{\lambda_1}^{\Lambda_1} \o{u}_{\mu_3}\r)^c \nn \\*
&\pm& \eps_{a b c} {1\over 2}
\l(                \o{u}^a_{\mu_1}
    {\cal D}_{\lambda_2}^{\Lambda_2} \o{d}^b_{\mu_2} -
                   \o{d}^a_{\mu_1}
    {\cal D}_{\lambda_2}^{\Lambda_2} \o{u}^b_{\mu_2} \r)
 \l({\cal D}_{\lambda_1}^{\Lambda_1} \o{u}_{\mu_3} \r)^c ,~
\label{eq:TwoQuarkDispNucleon}
\end{eqnarray}
and the following relation between spatial symmetry of
displacements and the symmetry of Dirac indices holds
\begin{eqnarray}
{\cal D}^{(1 + 2) \Lambda_2}_{\lambda_2}{\cal D}^{(3)
\Lambda_1}_{\lambda_1} \o N^{\rm (MA)}_{\mu_1\mu_2\mu_3}
\rightarrow \rm{ ~~A,~MA~Dirac~indices} \nonumber \\
{\cal D}^{(1 - 2) \Lambda_2}_{\lambda_2}{\cal D}^{(3)
\Lambda_1}_{\lambda_1} \o N^{\rm (MA)}_{\mu_1\mu_2\mu_3}
\rightarrow \rm{ ~~S,~MS~Dirac~indices }.
\end{eqnarray}
Because of symmetry, the operators of
Eq.~(\ref{eq:TwoQuarkDispNucleon}) with MA Dirac indices and
those with MS Dirac indices are identical.

Group theoretically, rotations of operators with one-link
displacements applied to two different quarks are the same as
those of two-link operators. Therefore the reduction to IRs is
exactly the same as for the two-link case. First use the
Clebsch-Gordan coefficients to obtain an IR for the product of
two displacements, and second use the Clebsch-Gordan coefficients
for the direct product of spatial and spinorial IRs to obtain
operators corresponding to overall IRs. The only additional
step is to determine the allowed symmetries of Dirac indices such
that the operator is antisymmetric under simultaneous exchange of
displacements, flavors, colors, and Dirac indices.

\section{Summary}
\label{sec:conclusions}

The constructions given in this paper provide a variety of
quasi-local and nonlocal three-quark operators for use as
zero-momentum baryon interpolating field operators in lattice QCD
simulations. All operators are categorized into the double-valued
IRs of the octahedral group $O^D_h$, they have definite parities
and they are gauge invariant. Operators correspond as closely as
possible to the continuum $|Jm\rangle$ IRs and they should be
useful for spectroscopy and for applications that require baryons
with a definite spin projection.

Complete sets of quasi-local operators are presented in
Section~\ref{sec:local} for each baryon. These quasi-local
constructions provide templates for the Dirac indices that should
be used to construct nonlocal operators. Nonlocal operators are
developed in Section~\ref{sec:nonlocal} based on adding
combinations of one-link displacements to one or more quarks. By
use of the building blocks given in this paper, a variety of
additional operators can be constructed by 1) using the
Clebsch-Gordan series to form overall IRs of the spatial
distribution, and 2) combining the spatial IRs with IRs of Dirac
indices to form operators corresponding to overall IRs.
Identification of the correct symmetry of Dirac indices is
straightforward when space-symmetric or space-antisymmetric
combinations of displacements are used.

Reference~\cite{Sato04} has demonstrated numerically that our
quasi-local and one-link operators are orthogonal in the sense of
Eq.~(\ref{eq:orthogonality}), i.e., a correlation function
vanishes if sink and source operators belong to different IRs
and rows.  For calculations of baryon masses, one should select
source operators within a fixed IR and row from the various
tables. Using operators from different embeddings of the IR
and row, matrices of correlation functions may be calculated and
mass spectra extracted. Correlation matrices can be made
hermitian by including a $\gamma_4$ matrix for each quark in the
source operator. Operators from our tables have the form $\o
B^{\Lambda,k}_\lambda=c^{(\Lambda,\lambda,k)}_{\mu_1 \mu_2 \mu_3}
\o B_{\mu_1 \mu_2 \mu_3}$, where $\o B_{\mu_1 \mu_2 \mu_3}$ is an
elemental baryon operator and a summation over repeated indices
is understood. A hermitian matrix of correlation functions can be
calculated in following way,
\begin{eqnarray}
&&\hspace{-4mm} C'^{(\Lambda)}_{kk'}(t) =
\sum_x \l< 0\r| T B^{\Lambda,k}_\lambda({\bf x},t)
\o B^{\Lambda,k'}_\lambda(0) \l| 0\r> \Gamma_4 \nn \\*
&& =
\sum_x c^{(\Lambda,\lambda,k)*}_{\mu_1 \mu_2 \mu_3}
c^{(\Lambda,\lambda,k')}_{\mu_1' \mu_2' \mu_3'}
\l< 0\r| T B_{\mu_1 \mu_2 \mu_3}({\bf x},t)
\o B_{\nu_1 \nu_2 \nu_3}(0) \l| 0\r> \nn \\*
&& ~~~ \times
\gamma_{4 \nu_1 \mu_1'}
\gamma_{4 \nu_2 \mu_2'}
\gamma_{4 \nu_3 \mu_3'}.
\end{eqnarray}
Exploratory calculations for baryon spectra along this line have
been reported in Ref.~\cite{Basak04}.  For a given baryon, the
dimension of the matrix of correlation functions depends on the
choices that are made for spatial distributions (quasi-local,
one-link, two-link, \textit{etc.}) and the overall IR. For
nucleon operators with quasi-local and one-link displacements, 23
$G_{1g}$ operators, 28 $H_g$ operators, and 7 $G_{2g}$ operators
are available as shown in Table~\ref{table:NumOpsForSim}.
The numbers of operators in each IR and row can be extended
without limit by using two-link and three-link operators and by
using different choices of smearing.
\begin{table}[h]
\caption{Available numbers of nucleon operators with quasi-local
and
         with one-link
         displacements for $G_{1g}$ row 1, $G_{2g}$ row 1, and $H_g$ row 1.
     The numbers of ungerade operators are exactly the same.  The $T_1$
         $\o N^{\rm (MS)}$ operators are omitted because after a
         projection to zero total momentum they are equivalent
     to the $T_1$ $\o N^{\rm (MA)}$ operators.}
\begin{center}
\begin{tabular}{|lccccc|}
\hline
~~~Type & ~Eq.~ & Table &
$G_{1g}$ & $H_g$ & $G_{2g}$ \\
\hline
\hline
quasi-local & \ref{eq:uudMA-local} & \ref{table:LocalNucleon} &
3 & 1 & 0 \\
one-link $A_1$ & \ref{eq:uudMA_one-link} & \ref{table:LocalLambda} &
4 & 1 & 0 \\
one-link $E$   & \ref{eq:uudMA_one-link} &
\ref{table:One-LinkDWave}, \ref{table:LocalLambda} &
1 & 5 & 1 \\
one-link $T_1$ & \ref{eq:uudMA_one-link} &
\ref{table:One-LinkPWave}, \ref{table:LocalLambda} &
5 & 6 & 1 \\
one-link $A_1$ & \ref{eq:uudMS_one-link} & \ref{table:LocalSigma} &
4 & 3 & 0 \\
one-link $E$   & \ref{eq:uudMS_one-link} &
~\ref{table:One-LinkDWave}, \ref{table:LocalSigma}~ &
3 & 7 & 3 \\
one-link $T_1$ & \ref{eq:uudMS_one-link} &
\ref{table:One-LinkPWave}, \ref{table:LocalSigma} &
3 & 5 & 2 \\
\hline
\hline
~~~~total & & & 23 & 28 & 7 \\
\hline
\end{tabular}
\end{center}
\label{table:NumOpsForSim}
\end{table}


\newpage
\begin{acknowledgements}
 This work was supported by the
U.S. National Science Foundation through grants PHY-0354982 and
PHY-0300065, and by the U.S. Dept. of Energy under contracts
DE-AC05-84ER40150 and DE-FG02-93ER-40762.
\end{acknowledgements}


\appendix

\section{Dirac matrices}
\label{app:DiracBasis}

Various conventions for the Dirac matrices are useful.  Each
is related by a unitary transformation to the Dirac-Pauli representation
as follows,
\begin{equation}
\gamma_{\mu} = U \gamma_{\mu}^{\rm (DP)} U^{\dag},
\end{equation}
where
\begin{equation}
\gamma_j^{\rm (DP)} = \l(
\begin{array}{cc}
0 & -i\sigma_j \\
i\sigma_j & 0
\end{array}
\r), ~~\gamma_4^{\rm (DP)} = \l(
\begin{array}{cc}
1 &  0 \\
0 & -1
\end{array}
\r).
\end{equation}
The unitary matrix that generates the Weyl convention is
\begin{equation}
U^{(W)} = \frac{1 + \gamma^{\rm (DP)}_5 \gamma^{\rm (DP)}_4}{\sqrt{2}},
\end{equation}
and the unitary transformation that generates the DeGrand-Rossi
convention is
\begin{equation}
U^{\rm (DR)}=
\frac{-i \gamma^{\rm (DP)}_2+\gamma^{\rm (DP)}_1\gamma^{\rm (DP)}_3}
{\sqrt{2}}.
\end{equation}
A quark field expressed in terms of the Dirac-Pauli representation
may be re-expressed in the Degrand-Rossi convention, for example, by
\begin{equation}
q^{\rm (DP)}_{\mu} = \left( U^{\rm (DR) \dag }\right)_{\mu \nu}
q^{\rm (DR)}_{\nu}.
\label{eq:unitary-trans}
\end{equation}

In order to display the spin and $\rho$-parity of fields in a
transparent way, we employ spin $s$ subscripts and $\rho$ superscripts
in place of the four Dirac components $\mu$ = 1, 2, 3 and 4
of each quark field in
the Dirac-Pauli representation as shown in Table~\ref{table:Dirac_rho_s}.

This encoding of the Dirac indices is based on the
$SU(2)\ot SU(2)$ representation of the Dirac matrices, where the
first $SU(2)$ is generated by 2$\times$2 Pauli matrices for $\rho$-spin,
\begin{eqnarray}
\rho_1 = \left( \begin{array}{cc} 0 & 1 \\ 1 & 0  \end{array} \right) ,
\rho_2 = \left( \begin{array}{cc} 0 & -i \\ i & 0 \end{array} \right) ,
\rho_3 = \left( \begin{array}{cc} 1 & 0 \\ 0 & -1 \end{array} \right) ,
\end{eqnarray}
and the second $SU(2)$ is generated by the Pauli matrices for ordinary spin,
\begin{eqnarray}
\sigma_1 = \left( \begin{array}{cc} 0 & 1 \\ 1 & 0  \end{array} \right) ,
\sigma_2 = \left( \begin{array}{cc} 0 & -i \\ i & 0 \end{array} \right) ,
\sigma_3 = \left( \begin{array}{cc} 1 & 0 \\ 0 & -1 \end{array} \right) .
\end{eqnarray}
The $2\times 2$ identity matrices for $\rho$-spin and ordinary
spin are
\begin{eqnarray}
\rho_4 =
\sigma_4 = \left( \begin{array}{cc} 1 & 0 \\ 0 & 1  \end{array} \right).
\end{eqnarray}
In terms of these sets of 2$\times$2 matrices, the 4$\times$4
Dirac-Pauli matrices are expressed as direct products of
$\rho$-spin matrices and $s$-spin matrices as follows,
\begin{eqnarray}
I &=& \rho_4 \ot \sigma_4, \nonumber \\
\gamma_4 &=& \rho_3 \ot \sigma_4,  \nonumber \\
\gamma_5 &=& \rho_1 \ot \sigma_4,  \nonumber \\
\gamma_5 \gamma_4 &=& -i \rho_2 \ot \sigma_4, \nonumber \\
\gamma_5 \gamma_k &=& i\rho_3 \ot \sigma_k,  \nonumber \\
\gamma_k &=&  \rho_2 \ot \sigma_k,  \nonumber \\
\sigma_{4k} &=& \rho_1 \ot \sigma_k, \nonumber \\
\sigma_{kl} &=& -\eps_{klm} \rho_4 \ot \sigma_m,
\end{eqnarray}
where $k, l$, and $m$ take the values 1, 2 and 3.

\section{Symmetry of three Dirac fields}
\label{app:DiracSymmetry}


The Dirac indices categorized in each Young tableau in
Fig.~\ref{figure:DiracYoungTableaux} should be further reduced
into $G_1$ or $H$ IRs for the purpose of operator
construction. (There is no $G_2$ IR with three Dirac spinors.)
Decomposition of the Dirac index into $\rho$-parity and ordinary
two-component spin ($s$-spin) simplifies this process.

The S (totally symmetric), MS (mixed-symmetric), and MA (mixed-antisymmetric)
combinations of three $s$-spins are defined as follows.
\newcommand{\h}{\hspace{-1.0mm}}
\begin{eqnarray}
\hspace{-5mm}{\rm S}  &:&
\l| +\h +\h + \r>;
{\l| +\h +\h - \r>+\l| +\h -\h + \r>+\l| -\h +\h + \r> \over \sqrt{3}};
\nn \\*
& &{\l| +\h -\h - \r>+\l| -\h +\h - \r>+\l| -\h -\h + \r> \over \sqrt{3}};
\l| -\h -\h - \r> .
\label{eq:S}\\
\hspace{-5mm}{\rm MS} &:&
{1\over\sqrt{6}}\l(2\l| +\h +\h - \r>-\l| +\h -\h + \r> -\l|-\h +\h + \r> \r);
\nn \\*
&&-{1\over\sqrt{6}}\l(2\l|-\h-\h+\r>-\l| -\h +\h - \r> - \l| +\h -\h - \r> \r).
\label{eq:MS} \\
\hspace{-5mm}{\rm MA} &:& \hspace{-2mm}
{1\over \sqrt{2}} \l( \l| +\h -\h + \r> - \l| -\h +\h + \r> \r);
{1\over \sqrt{2}} \l( \l| +\h -\h - \r> - \l| -\h +\h - \r> \r).
\label{eq:MA}
\end{eqnarray}
The four states in Eq.~(\ref{eq:S}) are $\l| J,m \r>=$ $\l|
{3\over2}  {3\over2} \r>$, $\l| {3\over2}  {1\over2} \r>$, $\l|
{3\over2} -{1\over2} \r>$, and $\l| {3\over2} -{3\over2} \r>$
respectively. The two states in Eq.~(\ref{eq:MS}) are $\l|
{1\over2}  {1\over2} \r>$ and $\l| {1\over2} -{1\over2} \r>$
while the two states in Eq.~(\ref{eq:MA}) are also $\l|
{1\over2}  {1\over2} \r>$ and $\l| {1\over2} -{1\over2} \r>$. All
these states are orthogonal to one other. Because S states in
Eq.~(\ref{eq:S}) span total spin 3/2, they are the bases of an
$H$ IR (no matter which $\rho$'s are involved in making up the
Dirac indices).  The MS and MA states in Eq.~(\ref{eq:MS},
\ref{eq:MA}) span total spin 1/2, so they are the bases of $G_1$
IRs.

Products of three $\rho$-spins are categorized in exactly the
same way. The $\rho$-parity is given by the product
$\rho_1\rho_2\rho_3$. Direct products of states of three
$\rho$-spins and states of three $s$-spins are simple when they
are expressed in the bases of S, MS, and MA. For instance, ${\rm
MA}_\rho \ot {\rm S}_s$ with subscripts denoting $\rho$-spin and
$s$-spin describes eight states, four of which have positive
$\rho$-parity, and four of which have negative $\rho$-parity. The
four states of each $\rho$-parity span $H$ IRs because IRs of
$O^D$ are determined only by the $s$-spins.  The direct product
of ${\rm MA}_\rho \ot {\rm S}_s$, with $\l| {1\over2},-{1\over2}
\r>_\rho$ and $\l| {3\over2},{1\over2} \r>_s$ is written as
\[
{\l| +\h -\h - \r>_{\rho} - \l| -\h +\h - \r>_{\rho} \over \sqrt{2}}
\ot
{\l| +\h +\h - \r>_{s} +
 \l| +\h -\h + \r>_{s} +
 \l| -\h +\h + \r>_{s} \over \sqrt{3}}. \nn
\]
By evaluating the direct product one obtains
\[
\hspace{-1mm} {1\over\sqrt{6}} \l(
\l| ^{+--}_{++-} \r> +
\l| ^{+--}_{+-+} \r> +
\l| ^{+--}_{-++} \r> -
\l| ^{-+-}_{++-} \r> -
\l| ^{-+-}_{+-+} \r> -
\l| ^{-+-}_{-++} \r>  \r),
\]
where the notation $\l| ^{\rho_1 \rho_2 \rho_3}_{s_1 s_2 s_3}\r>$
is used. In terms of the $\l| \mu_1 \mu_2 \mu_3 \r>$ notation, it
becomes
\[
{1\over\sqrt{6}} \l(
\l| 134 \r> +
\l| 143 \r> +
\l| 233 \r> -
\l| 314 \r> -
\l| 323 \r> -
\l| 413 \r> \r),
\]
where the Dirac indices $\mu_i=1,2,3,4$ are defined in Dirac-Pauli
representation of Dirac $\gamma$ matrices.
The translation of $\mu$ to $(\rho,s)$ is given in
Table~\ref{table:Dirac_rho_s}.
It is clear that the obtained Dirac indices are antisymmetric under exchange
of first two labels but not totally antisymmetric.  Thus, we denote
${\rm MA}_\rho \ot {\rm S}_s = {\rm MA}_{\rm Dirac}$.  The
nucleon operator that follows from this
example is labeled as $H_g$, row 2 in Table~\ref{table:LocalNucleon}.

From such considerations one obtains
Table~\ref{table:Symm_Dirac_rho_s}, which provides the relations
of Dirac index symmetries (abbreviated as ``Dirac sym'' in the
table) to IRs of Dirac indices, and direct products of
$\rho$-spins and $s$-spins.
\renewcommand{\S}{{\rm S}}
\newcommand{\MS}{{\rm MS}}
\newcommand{\MA}{{\rm MA}}
\newcommand{\A}{{\rm A}}
\renewcommand{\D}{{\rm Dirac}}
\begin{table}[h]
\caption{Relation among Dirac spin symmetry, IR, and the direct
product of
         $\rho$-spins and $s$-spins.}
\begin{center}
\label{table:Symm_Dirac_rho_s}
\begin{tabular}{|clcc|}
\hline
Dirac sym & IR & emb & $\rho \ot s$ \\
\hline \hline
$\S_\D$  & $G_1$ &  1  & $\MA_\rho \ot \MA_s \op \MS_\rho \ot \MS_s$ \\
         & $ H $ & 1,2 & $\S_\rho \ot \S_s$ \\
$\MS_\D$ & $G_1$ & 1,2 & $\S_\rho \ot \MS_s$ \\
         &       &  3  & $\MA_\rho \ot \MA_s \om \MS_\rho \ot \MS_s$ \\
         & $ H $ &  1  & $\MS_\rho \ot \S_s$ \\
$\MA_\D$ & $G_1$ & 1,2 & $\S_\rho \ot \MA_s$ \\
         &       &  3  & $\MA_\rho \ot \MS_s \op \MS_\rho \ot \MA_s$ \\
         & $ H $ &  1  & $\MA_\rho \ot \S_s$ \\
$\A_\D$  & $G_1$ & 1,2 & $\MA_\rho \ot \MS_s \om \MS_\rho \ot \MA_s$ \\
\hline
\end{tabular}
\end{center}
\end{table}
Note that $\MA_\rho \ot \MA_s$ and $\MS_\rho \ot \MS_s$ both have
a mixture of $\S_\D$ and $\MS_\D$.  One can easily see that
addition of a state from $\MA_\rho \ot \MA_s$, say $G_{1g}$, row
1, and a state from $\MS_\rho \ot \MS_s$ of the same $G_{1g}$,
row 1 yields a pure $\S_\D$ state. The subtraction of the states
yields a pure $\MS_\D$ state.  Similarly, $\MA_\rho \ot \MS_s$
and $\MS_\rho \ot \MA_s$ have a mixture of $\MA_\D$ and $\A_\D$.
A pure $\MA_\D$ is obtained by addition of states from $\MA_\rho
\ot \MS_s$ and $\MS_\rho \ot \MA_s$ and a pure $\A_\D$ is
obtained by subtraction of states from $\MA_\rho \ot \MS_s$ and
$\MS_\rho \ot \MA_s$.  The third column of
Table~\ref{table:Symm_Dirac_rho_s} shows an embedding that has a
connection to Table~\ref{table:sixtyfour1} in a self-explanatory
way.

\begin{table}[h]
\caption{IRs of S, MS combinations of Dirac indices for three
quark states
         in Dirac-Pauli representation.  The first entry of MS table reads
         $2\o q^{f_1}_1 \o q^{f_2}_1 \o q^{f_3}_2 -
     \o q^{f_1}_1 \o q^{f_2}_2 \o q^{f_3}_1 -
     \o q^{f_1}_2 \o q^{f_2}_1 \o q^{f_3}_1$
     for the $G_{1g}$, embedding 1, row 1 local operator.}
\label{table:sixtyfour1}
\renewcommand{\arraystretch}{1.0}
\begin{center}
{\scriptsize
\begin{tabular}{|lccc|}
\hline
$\Lambda$ & $k$ & $\lambda$  &
S Dirac indices $\mu_1 \mu_2 \mu_3$ \\
\hline
\hline
  $G_{1g}$ & 1 & 1 & $-2(332+323+233)+341+431+314+413+134+143$ \\
           &   & 2 & $2(144+414+441)-234-342-423-243-324-432$ \\
  $G_{1u}$ & 1 & 1 & $2(114+141+411)-123-213-132-231-321-312$ \\
           &   & 2 & $-2(223+232+322)+214+124+241+142+421+412$ \\
  $H_g$    & 1 & 1 & $111$ \\
           &   & 2 & $112+121+211$ \\
           &   & 3 & $122+212+221$ \\
           &   & 4 & $222$ \\
           & 2 & 1 & $133+313+331$ \\
           &   & 2 & $233+323+332+134+341+413+143+431+314$ \\
           &   & 3 & $144+414+441+234+342+423+243+432+324$ \\
           &   & 4 & $244+424+442$ \\
  $H_u$    & 1 & 1 & $113+131+311$ \\
           &   & 2 & $411+141+114+312+123+231+321+213+132$ \\
           &   & 3 & $322+232+223+412+124+241+421+214+142$ \\
           &   & 4 & $224+242+422$ \\
           & 2 & 1 & $333$ \\
           &   & 2 & $334+343+433$ \\
           &   & 3 & $344+434+443$ \\
           &   & 4 & $444$ \\
\hline
\hline
$\Lambda$ & $k$ & $\lambda$  &
MS Dirac indices $\mu_1 \mu_2 \mu_3$ \\
\hline
\hline
  $G_{1g}$ & 1 & 1 & $2(112)-121-211$ \\
           &   & 2 & $-2(221)+212+122$ \\
           & 2 & 1 & $2(332+314+134)-341-323-143-431-413-233$ \\
           &   & 2 & $-2(441+423+243)+432+414+234+342+324+144$ \\
           & 3 & 1 & $-2(332+413+143)+323+233+134+314+341+431$ \\
           &   & 2 & $2(441+324+234)-414-144-243-423-432-342$ \\
  $G_{1u}$ & 1 & 1 & $2(114+132+312)-123-141-321-213-231-411$ \\
           &   & 2 & $-2(223+241+421)+214+232+412+124+142+322$ \\
           & 2 & 1 & $2(334)-343-433$  \\
           &   & 2 & $-2(443)+434+344$ \\
           & 3 & 1 & $2(114+231+321)-141-411-312-132-123-213$ \\
           &   & 2 & $-2(223+142+412)+232+322+421+241+214+124$ \\
  $H_g$    & 1 & 1 & $-2(331)+313+133$ \\
           &   & 2 & $-2(332+341+431)+314+134+323+143+413+233$ \\
           &   & 3 & $-2(342+432+441)+324+144+414+234+423+243$ \\
           &   & 4 & $-2(442)+424+244$ \\
  $H_u$    & 1 & 1 & $2(113)-131-311$ \\
           &   & 2 & $2(114+123+213)-132-312-141-321-231-411$ \\
           &   & 3 & $2(124+214+223)-142-322-232-412-241-421$ \\
           &   & 4 & $2(224)-242-422$ \\
\hline
\end{tabular}
}
\end{center}
\end{table}

\begin{table}[h]
\caption{IRs of MA, A combinations of Dirac indices for three-quark states
         in Dirac-Pauli representation.
     The caption in Table~\ref{table:sixtyfour1} describes how to read this
     table.}
\label{table:sixtyfour2}
\renewcommand{\arraystretch}{1.0}
\begin{center}
{\footnotesize
\begin{tabular}{|lccc|}
\hline
$\Lambda$ & $k$ & $\lambda$  &
MA Dirac indices $\mu_1 \mu_2 \mu_3$ \\
\hline
\hline
   $G_{1g}$ & 1 & 1 & $121-211$ \\
            &   & 2 & $122-212$ \\
            & 2 & 1 & $143-233+323-413+341-431$ \\
            &   & 2 & $144-234+324-414+342-432$ \\
            & 3 & 1 & $-233+323+134-314-341+431$ \\
            &   & 2 & $-414+144-243+423+432-342$ \\
   $G_{1u}$ & 1 & 1 & $123-213+141-231+321-411$ \\
            &   & 2 & $124-214+142-232+322-412$ \\
            & 2 & 1 & $343-433$ \\
            &   & 2 & $344-434$ \\
            & 3 & 1 & $-141+411-312+132+123-213$ \\
            &   & 2 & $232-322+421-241-214+124$ \\
   $H_g$    & 1 & 1 & $133-313$ \\
            &   & 2 & $134-314+143-323+233-413$ \\
            &   & 3 & $144-324+234-414+243-423$ \\
            &   & 4 & $244-424$ \\
   $H_u$    & 1 & 1 & $131-311$ \\
            &   & 2 & $132-312+141-321+231-411$ \\
            &   & 3 & $142-322+232-412+241-421$ \\
            &   & 4 & $242-422$ \\
\hline
\hline
$\Lambda$ & $k$ & $\lambda$  &
A Dirac indices $\mu_1 \mu_2 \mu_3$ \\
\hline
\hline
   $G_{1g}$ & 1 & 1 & $134-314+341-431+413-143$ \\
            &   & 2 & $234-324+342-432+423-243$ \\
   $G_{1u}$ & 1 & 1 & $-123+213-231+321-312+132$ \\
            &   & 2 & $-124+214-241+421-412+142$ \\
\hline
\end{tabular}
}
\end{center}
\end{table}
Explicit combinations of Dirac indices $\mu_1\mu_2\mu_3$ are
given in Tables~\ref{table:sixtyfour1}
and~\ref{table:sixtyfour2}.  Table~\ref{table:sixtyfour1} contains
all S and MS combinations of three Dirac indices, assigning each
to an IR ( $G_{1g/u}$ or $H_{g/u}$), embedding, and row.
Table~\ref{table:sixtyfour2} contains all MA and A combinations
of three Dirac indices in a similar way.

\section{Relations of $N_{\mu_1\mu_2\mu_3}$ to
         commonly used nucleon operators}
\label{app:chi1_chi2}

Various groups have performed lattice simulations using
the following two interpolating fields for a nucleon:
\begin{eqnarray}
\chi^{1/2}_1 &=& \l( u^T C \gamma_5 d \r) u, \\
\chi^{1/2}_2 &=& \l( u^T C d \r) \gamma_5 u,
\end{eqnarray}
where spacetime arguments are omitted.  Matrix $C$ is a
charge-conjugation operator, defined by $C=\gamma_4 \gamma_2$.
Each of these four-component operators corresponds to a $G_1$
IR and may be written in terms of $\o
\Psi^{\Lambda,k}_{S,S_z}$. Positive and negative $\rho$-parity
parts of $\chi^{1/2}_1$ are projected in Dirac-Pauli
representation as follows,
\begin{eqnarray}
\hspace{-2mm}
{1+\gamma_4 \over 2} \chi^{1/2}_1 \hspace{-2mm} &=& \hspace{-2mm} \l(
\begin{tabular}{c}
$-N_{121}-N_{341}$ \\
$-N_{122}-N_{342}$
\end{tabular}
\r) \nn \\*
\hspace{-2mm} {1-\gamma_4 \over 2} \chi^{1/2}_1 \hspace{-2mm}
&=&
\hspace{-2mm} \l(
\begin{tabular}{c}
$-N_{123}-N_{343}$ \\
$-N_{124}-N_{344}$
\end{tabular}
\r) \nn \\*
\end{eqnarray}
The upper component corresponds to $S_z = +1/2$.
Similarly $\chi^{1/2}_2$  can be projected to operators of definite
$\rho$-parity,
\begin{eqnarray}
{1+\gamma_4 \over 2} \chi^{1/2}_2 &=& \l(
\begin{tabular}{c}
$N_{143}-N_{233}$ \\
$N_{144}-N_{234}$
\end{tabular}
\r) \nn \\*
{1-\gamma_4 \over 2} \chi^{1/2}_2 &=& \l(
\begin{tabular}{c}
$N_{141}-N_{231}$ \\
$N_{142}-N_{232}$
\end{tabular}
\r) \nn \\* 
.
\end{eqnarray}
These results show how the components of $\chi_1^{1/2}$ and
$\chi_2^{1/2}$ are related to operators defined in this paper.

\section{Clebsch-Gordan coefficients for the double octahedral group}
\label{app:Altmann}

The Clebsch-Gordan formula shows how an IR operator $\o{\cal
O}^{\Lambda}_\lambda$  may be built from linear combinations of
direct products of other IR operators, $\o{\cal
O}^{\Lambda_1}_{\lambda_1}$ and $ \o{\cal
O}^{\Lambda_2}_{\lambda_2}$,
\begin{eqnarray}
\o{\cal O}^{\Lambda}_\lambda = \sum_{\lambda_1,\lambda_2}
C \l( \hspace{-0.5mm}
{\footnotesize
\renewcommand{\arraystretch}{0.41}
\begin{tabular}{ccc}
$\Lambda$ & $\Lambda_1$ & $\Lambda_2$ \\
$\lambda$ & $\lambda_1$ & $\lambda_2$
\end{tabular} }
\hspace{-0.8mm} \r)
\o{\cal O}^{\Lambda_1}_{\lambda_1} \o{\cal O}^{\Lambda_2}_{\lambda_2},
\label{eq:CG_operator}
\end{eqnarray}
where $\Lambda$ $(\Lambda_i)$ and $\lambda$ $(\lambda_i)$ denote
IR and row, respectively. The notation $\o{\cal
O}^\Lambda_\lambda$ can refer to
$\hat{A}_1,\hat{A}_2^\lambda,\hat{E}^\lambda,\hat{T}_1^\lambda,
\hat{T}_2^\lambda, \o \Psi ^{G_{1g}}_\lambda,\o
\Psi^{G_{1u}}_\lambda, \o \Psi ^{H_g}_\lambda$, or $\o \Psi
^{H_u}_\lambda$. See the comments in the paragraph above
Eq.~(\ref{eq:CG_coefficient}) for our phase convention for the
coefficients.

A complete set of Clebsch-Gordan coefficients for the octahedral
group using the basis vectors of
Tables~\ref{table:lattice_harmonics} and \ref{tab:spin_harmonics}
is given here and in Ref.~\cite{IkuroThesis}. The version of this
paper that was submitted for publication was shortened by omitting
all but a selected subset of Clebsch-Gordan coefficients.

In each Clebsch-Gordan table, the resultant IR appearing on the left side
of Eq.~(\ref{eq:CG_operator}) is listed in the top row, and the
two IRs appearing on the right side of
Eq.~(\ref{eq:CG_operator}) are listed in the left column.
Table~\ref{table:HowToReadCG}
explains how to read the coefficients in the Clebsch-Gordan
tables in this appendix.

\begin{table}[h]
\caption{Description for tables of Clebsch-Gordan coefficients.
         Squares of coefficients are listed together with their overall sign.}
\begin{tabular}{|c|c|}
\hline \hline
$\o{\cal O}^{\Lambda_1} \ot \o{\cal O}^{\Lambda_2}$
&
$\o{\cal O}^\Lambda_\lambda$ \\
\hline $\o{\cal O}^{\Lambda_1}_{\lambda_1} \o{\cal
O}^{\Lambda_2}_{\lambda_2}$
&
$sgn\l[ C \l( \hspace{-0.5mm}
{\footnotesize
\begin{array}{ccc}
\Lambda & \Lambda_1 & \Lambda_2 \\
\lambda & \lambda_1 & \lambda_2
\end{array} }
\hspace{-0.8mm} \r) \r]
\l| C \l( \hspace{-0.5mm}
{\footnotesize
\begin{array}{ccc}
\Lambda & \Lambda_1 & \Lambda_2 \\
\lambda & \lambda_1 & \lambda_2
\end{array} }
\hspace{-0.8mm} \r) \r|^2 $ \\
\hline \hline
\end{tabular}
\label{table:HowToReadCG}
\end{table}
\renewcommand{\ot}{\!\!\otimes\!}
\begin{tabular}{|l|rrrrrr|}
\hline \hline $E \ot T_1$ & $T_1^1$
&$T_1^2$&$T_1^3$&$T_2^1$&$T_2^2$ &$T_2^3$ \\
\hline
$E^1T_1^1$  &  1/4 & 0  &  0  & 3/4 &  0  &  0  \\
$E^1T_1^2$  &  0   &-1  &  0  &  0  &  0  &  0  \\
$E^1T_1^3$  &  0   & 0  & 1/4 &  0  &  0  &-3/4 \\
$E^2T_1^1$  &  0   & 0  & 3/4 &  0  &  0  & 1/4 \\
$E^2T_1^2$  &  0   & 0  &  0  &  0  & -1  &  0  \\
$E^2T_1^3$  &  3/4 & 0  &  0  &-1/4 &  0  &  0  \\
\hline \hline
\end{tabular}
\begin{tabular}{|l|rrrr|}
\hline
\hline
$E\ot G_1$  &$H^1$& $H^2$& $H^3$ & $H^4$ \\
\hline
$E^1G_1^1$  &  0  &-1  & 0  & 0  \\
$E^1G_1^2$  &  0  & 0  & 1  & 0  \\
$E^2G_1^1$  &  0  & 0  & 0  &-1  \\
$E^2G_1^2$  &  1  & 0  & 0  & 0  \\
\hline
\hline
\end{tabular}
\begin{tabular}{|l|rrrrrr|}
\hline
\hline
$T_1 \ot G_1$ & $G_1^1$ &$G_1^2$&$H^1$&$H^2$&$ H^3$&$H^4$ \\
\hline
$T_1^1G_1^1$  &  0  &  0  &  1  &  0  &  0  &  0  \\
$T_1^1G_1^2$  & 2/3 &  0  &  0  & 1/3 &  0  &  0  \\
$T_1^2G_1^1$  &-1/3 &  0  &  0  & 2/3 &  0  &  0  \\
$T_1^2G_1^2$  &  0  & 1/3 &  0  &  0  & 2/3 &  0  \\
$T_1^3G_1^1$  &  0  &-2/3 &  0  &  0  & 1/3 &  0  \\
$T_1^3G_1^2$  &  0  &  0  &  0  &  0  &  0  &  1  \\
\hline
\hline
\end{tabular}
\begin{tabular}{|l|rrrrrrrr|}
\hline \hline $E \ot H$ & $G_1^1$&$G_1^2$&$G_2^1$&$G_2^2$&
$H^1$     & $H^2$ & $H^3$ & $H^4$ \\
\hline
$E^1H^1$  &  0  &   0  &   0  & -1/2 &  1/2 &   0  &   0  &   0  \\
$E^1H^2$  & 1/2 &   0  &   0  &   0  &   0  & -1/2 &   0  &   0  \\
$E^1H^3$  &  0  & -1/2 &   0  &   0  &   0  &   0  & -1/2 &   0  \\
$E^1H^4$  &  0  &   0  &  1/2 &   0  &   0  &   0  &   0  &  1/2 \\
$E^2H^1$  &  0  & -1/2 &   0  &   0  &   0  &   0  &  1/2 &   0  \\
$E^2H^2$  &  0  &   0  & -1/2 &   0  &   0  &   0  &   0  &  1/2 \\
$E^2H^3$  &  0  &   0  &   0  &  1/2 &  1/2 &   0  &   0  &   0  \\
$E^2H^4$  & 1/2 &   0  &   0  &   0  &   0  &  1/2 &   0  &   0  \\
\hline \hline
\end{tabular}
\begin{tabular}{|l|rr|}
\hline
$A_2\ot E$ & $E^1$& $E^2$ \\
\hline
$A_2 E^1$  &  0  &  -1  \\
$A_2 E^2$  &  1  &   0  \\
\hline
\end{tabular}
\begin{tabular}{|l|rrr|}
\hline
$A_2\ot T_1$ &$T_2^1$& $T_2^2$& $T_2^3$ \\
\hline
$A_2 T_1^1$  &  0  &  0  &-1  \\
$A_2 T_1^2$  &  0  & -1  & 0  \\
$A_2 T_1^3$  &  1  &  0  & 0  \\
\hline
\end{tabular}
\begin{tabular}{|l|rrr|}
\hline
$A_2\ot T_2$ &$T_1^1$& $T_1^2$& $T_1^3$ \\
\hline
$A_2 T_2^1$ &  0  &  0  & -1 \\
$A_2 T_2^2$ &  0  &  1  &  0 \\
$A_2 T_2^3$ &  1  &  0  &  0 \\
\hline
\end{tabular}
\begin{tabular}{|l|rr|}
\hline
$A_2\ot G_1$ & $G_2^1$& $G_2^2$ \\
\hline
$A_2 G_1^1$  &  1  &  0  \\
$A_2 G_1^2$  &  0  &  1  \\
\hline
\end{tabular}
\begin{tabular}{|l|rr|}
\hline
$A_2\ot G_2$ & $G_1^1$& $G_1^2$ \\
\hline
$A_2 G_2^1$  &  1  &  0  \\
$A_2 G_2^2$  &  0  &  1  \\
\hline
\end{tabular}
\begin{tabular}{|l|rrrr|}
\hline
$A_2\ot H$ &$H^1$& $H^2$& $H^3$ &$H^4$ \\
\hline
$A_2 H^1$  &  0  &  0  &  -1 & 0  \\
$A_2 H^2$  &  0  &  0  &   0 & 1  \\
$A_2 H^3$  &  1  &  0  &   0 & 0  \\
$A_2 H^4$  &  0  & -1  &   0 & 0  \\
\hline
\end{tabular}
\begin{tabular}{|l|rrrr|}
\hline
$E\ot E$  &$A_1$&$A_2$ & $E^1$& $E^2$ \\
\hline
$E^1E^1$  & 1/2 &   0  &  1/2 &    0  \\
$E^1E^2$  &  0  &  1/2 &   0  & -1/2 \\
$E^2E^1$  &  0  & -1/2 &   0  & -1/2 \\
$E^2E^2$  & 1/2 &   0  & -1/2 &    0  \\
\hline
\end{tabular}
%
\begin{tabular}{|l|rrrr|}
\hline
$E\ot G_2$  &$H^1$& $H^2$& $H^3$ & $H^4$ \\
\hline
$E^1G_2^1$  &  0  & 0  &  0  &-1  \\
$E^1G_2^2$  &  1  & 0  &  0  & 0  \\
$E^2G_2^1$  &  0  & 1  &  0  & 0  \\
$E^2G_2^2$  &  0  & 0  & -1  & 0  \\
\hline
\end{tabular}
%
%
\begin{tabular}{|l|rrrrrr|}
\hline $E \ot T_2$ & $T_1^1$ &$T_1^2$&$ T_1^3$&$ T_2^1$&$ T_2^2$
& $T_2^3$ \\
\hline
$E^1T_2^1$  &  3/4 &  0  &   0  &  1/4 & 0  &   0  \\
$E^1T_2^2$  &   0  &  0  &   0  &   0  &-1  &   0  \\
$E^1T_2^3$  &   0  &  0  & -3/4 &   0  & 0  &  1/4 \\
$E^2T_2^1$  &   0  &  0  & -1/4 &   0  & 0  & -3/4 \\
$E^2T_2^1$  &   0  & -1  &   0  &   0  & 0  &   0  \\
$E^2T_2^1$  &  1/4 &  0  &   0  & -3/4 & 0  &   0  \\
\hline
\end{tabular}
%
\begin{tabular}{|l|rrrrrrrrr|}
\hline $T_1 \ot T_1$  & $A_1$ &$E^1$&$E^2$&$T_1^1$&$
T_1^2$&$T_1^3$
&$ T_2^1$&$ T_2^2$&$T_2^3$ \\
\hline
$T_1^1 T_1^1$  &   0  & 0  & 1/2 &  0  &  0  &  0  & 0  & 1/2 &  0  \\
$T_1^1 T_1^2$  &   0  & 0  &  0  & 1/2 &  0  &  0  &1/2 &  0  &  0  \\
$T_1^1 T_1^3$  &  1/3 & 1/6&  0  &  0  & 1/2 &  0  & 0  &  0  &  0  \\
$T_1^2 T_1^1$  &   0  & 0  &  0  &-1/2 &  0  &  0  &1/2 &  0  &  0  \\
$T_1^2 T_1^2$  & -1/3 & 2/3&  0  &  0  &  0  &  0  & 0  &  0  &  0  \\
$T_1^2 T_1^3$  &   0  & 0  &  0  &  0  &  0  & 1/2 & 0  &  0  & 1/2 \\
$T_1^3 T_1^1$  &  1/3 & 1/6&  0  &  0  &-1/2 &  0  & 0  &  0  &  0  \\
$T_1^3 T_1^2$  &   0  & 0  &  0  &  0  &  0  &-1/2 & 0  &  0  & 1/2 \\
$T_1^3 T_1^3$  &   0  & 0  & 1/2 &  0  &  0  &  0  & 0  &-1/2 &  0  \\
\hline
\end{tabular}
%
{\tabcolsep=0.1mm \small
\begin{tabular}{|l|rrrrrrrrrrrr|}
\hline \hline $T_{\! 1} \ot H$&
$G_1^1$&$G_1^2$&$G_2^1$&$G_2^2$&$H^1$&$H^2$&
$ H^3$&$H^4$ &$H^1$&$H^2$&$ H^3$&$H^4$ \\
\hline
$T_1^1H^1$ &  0  & 0  &1/6 & 0  & 0  &  0  &  0  & 0  & 0  &  0  & 0  &-5/6 \\
$T_1^1H^2$ &  0  & 0  & 0  &-1/2&-2/5&  0  &  0  & 0  &1/10&  0  & 0  & 0   \\
$T_1^1H^3$ & 1/6 & 0  & 0  & 0  & 0  &-8/15&  0  & 0  & 0  &-3/10& 0  & 0   \\
$T_1^1H^4$ &  0  &1/2 & 0  & 0  & 0  &  0  & -2/5& 0  & 0  &  0  &1/10& 0   \\
$T_1^2H^1$ &  0  & 0  & 0  &-1/3&3/5 &  0  &  0  & 0  &1/15&  0  & 0  & 0   \\
$T_1^2H^2$ &-1/3 & 0  & 0  & 0  & 0  &1/15 &  0  & 0  & 0  & -3/5& 0  & 0   \\
$T_1^2H^3$ &  0  &-1/3& 0  & 0  & 0  &  0  &-1/15& 0  & 0  &  0  &3/5 & 0   \\
$T_1^2H^4$ &  0  & 0  &-1/3& 0  & 0  &  0  &  0  &-3/5& 0  &  0  & 0  &-1/15\\
$T_1^3H^1$ & 1/2 & 0  & 0  & 0  & 0  & 2/5 &  0  & 0  & 0  &-1/10& 0  & 0   \\
$T_1^3H^2$ &  0  &1/6 & 0  & 0  & 0  &  0  & 8/15& 0  & 0  &  0  &3/10& 0   \\
$T_1^3H^3$ &  0  & 0  &-1/2& 0  & 0  &  0  &  0  &2/5 & 0  &  0  & 0  &-1/10\\
$T_1^3H^4$ &  0  & 0  & 0  &1/6 & 0  &  0  &  0  & 0  &5/6 &  0  & 0  & 0   \\
\hline \hline
\end{tabular}
}
\begin{tabular}{|l|rrrrrrrrr|}
\hline $T_1 \ot T_2$& $A_2$ &$E^1$&$E^2$&$T_1^1$&$ T_1^2$&$T_1^3$
&$ T_2^1$&$ T_2^2$&$T_2^3$ \\
\hline
$T_1^1T_2^1$ & 1/3 &  0  &1/6 &  0  &  0  & 0  & 0  &-1/2&  0  \\
$T_1^1T_2^2$ & 0   &  0  & 0  &  0  &  0  & 1/2& 0  & 0  & 1/2 \\
$T_1^1T_2^3$ &  0  & 1/2 & 0  &  0  &-1/2 & 0  & 0  & 0  &  0  \\
$T_1^2T_2^1$ &  0  &  0  & 0  & 1/2 &  0  & 0  & 1/2& 0  &  0  \\
$T_1^2T_2^2$ & 1/3 &  0  &-2/3&  0  &  0  & 0  & 0  & 0  &  0  \\
$T_1^2T_2^3$ &  0  &  0  & 0  &  0  &  0  & 1/2& 0  & 0  &-1/2 \\
$T_1^3T_2^1$ &  0  &-1/2 & 0  &  0  &-1/2 & 0  & 0  & 0  &  0  \\
$T_1^3T_2^2$ &  0  &  0  & 0  &-1/2 &  0  & 0  & 1/2& 0  &  0  \\
$T_1^3T_2^3$ &-1/3 &  0  &-1/6&  0  &  0  & 0  & 0  &-1/2&  0  \\
\hline
\end{tabular}
%
\begin{tabular}{|l|rrrrrr|}
\hline
$T_1 \ot G_2$& $G_2^1$ &$G_2^2$&$H^1$&$H^2$&$ H^3$&$H^4$ \\
\hline
$T_1^1G_2^1$ &  0  &  0  &  0  & 0  &-1  & 0  \\
$T_1^1G_2^2$ & 2/3 &  0  &  0  & 0  & 0  &1/3 \\
$T_1^2G_2^1$ &-1/3 &  0  &  0  & 0  & 0  &2/3 \\
$T_1^2G_2^2$ &  0  & 1/3 & 2/3 & 0  & 0  & 0  \\
$T_1^3G_2^1$ &  0  &-2/3 & 1/3 & 0  & 0  & 0  \\
$T_1^3G_2^2$ &  0  &  0  &  0  &-1  & 0  & 0  \\
\hline
\end{tabular}
%
\begin{tabular}{|l|rrrrrrrrr|}
\hline $T_2 \ot T_2$& $A_1$ &$E^1$&$E^2$&$T_1^1$&$ T_1^2$&$T_1^3$
&$ T_2^1$&$ T_2^2$&$T_2^3$ \\
\hline
$T_2^1T_2^1$ &  0  &  0  &-1/2 &  0  & 0  & 0  & 0  &-1/2&  0  \\
$T_2^1T_2^2$ &  0  &  0  &  0  &  0  & 0  &-1/2& 0  & 0  &-1/2 \\
$T_2^1T_2^3$ & 1/3 & 1/6 &  0  &  0  &-1/2& 0  & 0  & 0  &  0  \\
$T_2^2T_2^1$ &  0  &  0  &  0  &  0  & 0  & 1/2& 0  & 0  &-1/2 \\
$T_2^2T_2^2$ & 1/3 &-2/3 &  0  &  0  & 0  & 0  & 0  & 0  &  0  \\
$T_2^2T_2^3$ &  0  &  0  &  0  & 1/2 & 0  & 0  &1/2 & 0  &  0  \\
$T_2^3T_2^1$ & 1/3 & 1/6 &  0  &  0  &1/2 & 0  & 0  & 0  &  0  \\
$T_2^3T_2^2$ &  0  &  0  &  0  &-1/2 & 0  & 0  &1/2 & 0  &  0  \\
$T_2^3T_2^3$ &  0  &  0  &-1/2 &  0  & 0  & 0  & 0  &1/2 &  0  \\
\hline
\end{tabular}
\begin{tabular}{|l|rrrrrr|}
\hline
$T_2 \ot G_1$& $G_2^1$ &$G_2^2$&$H^1$&$H^2$&$ H^3$&$H^4$ \\
\hline
$T_2^1G_1^1$ &  0  &-2/3 & 1/3 &  0  & 0  &  0   \\
$T_2^1G_1^2$ &  0  &  0  &  0  & -1  & 0  &  0   \\
$T_2^2G_1^1$ & 1/3 &  0  &  0  &  0  & 0  &-2/3  \\
$T_2^2G_1^2$ &  0  &-1/3 &-2/3 &  0  & 0  &  0   \\
$T_2^3G_1^1$ &  0  &  0  &  0  &  0  & 1  &  0   \\
$T_2^3G_1^2$ &-2/3 &  0  &  0  &  0  & 0  &-1/3  \\
\hline
\end{tabular}
\begin{tabular}{|l|rrrrrr|}
\hline
$T_2 \ot G_2$& $G_1^1$ &$G_1^2$&$H^1$&$H^2$&$ H^3$&$H^4$ \\
\hline
$T_2^1G_2^1$ &  0  &-2/3& 0  & 0  &-1/3& 0  \\
$T_2^1G_2^2$ &  0  & 0  & 0  & 0  & 0  &-1  \\
$T_2^2G_2^1$ & 1/3 & 0  & 0  &2/3 & 0  & 0  \\
$T_2^2G_2^2$ &  0  &-1/3& 0  & 0  & 2/3& 0  \\
$T_2^3G_2^1$ &  0  & 0  & 1  & 0  & 0  & 0  \\
$T_2^3G_2^2$ &-2/3 & 0  & 0  &1/3 & 0  & 0  \\
\hline
\end{tabular}
{\tabcolsep=0.2mm
\begin{tabular}{|l|rrrrrrrrrrrr|}
\hline $T_2 \ot H$& $G_1^1$&$G_1^2$&$G_2^1$&$G_2^2$&$H^1$&$H^2$&
$ H^3$&$H^4$ &$H^1$&$H^2$&$ H^3$&$H^4$ \\
\hline
$T_2^1H^1$ & 0  & 0  &1/2 & 0  & 0   & 0  & 0  &  0  &  0  &  0  & 0  &1/2 \\
$T_2^1H^2$ & 0  & 0  & 0  &1/6 &2/3  & 0  & 0  &  0  & 1/6 &  0  & 0  & 0  \\
$T_2^1H^3$ &1/2 & 0  & 0  & 0  & 0   & 0  & 0  &  0  &  0  &-1/2 & 0  & 0  \\
$T_2^1H^4$ & 0  &-1/6& 0  & 0  & 0   & 0  &-2/3&  0  &  0  &  0  & 1/6& 0  \\
$T_2^2H^1$ & 0  &-1/3& 0  & 0  & 0   & 0  & 1/3&  0  &  0  &  0  & 1/3& 0  \\
$T_2^2H^2$ & 0  & 0  & 1/3& 0  & 0   & 0  & 0  & 1/3 &  0  &  0  & 0  &-1/3\\
$T_2^2H^3$ & 0  & 0  & 0  & 1/3& -1/3& 0  & 0  &  0  & 1/3 &  0  & 0  & 0  \\
$T_2^2H^4$ &-1/3& 0  & 0  & 0  & 0   &-1/3& 0  &  0  &  0  & -1/3& 0  & 0  \\
$T_2^3H^1$ &1/6 & 0  & 0  & 0  & 0   &-2/3& 0  &  0  &  0  & 1/6 & 0  & 0  \\
$T_2^3H^2$ & 0  &-1/2& 0  & 0  & 0   & 0  & 0  &  0  &  0  &  0  &-1/2& 0  \\
$T_2^3H^3$ & 0  & 0  &-1/6& 0  & 0   & 0  & 0  & 2/3 &  0  &  0  & 0  &1/6 \\
$T_2^3H^4$ & 0  & 0  & 0  &-1/2& 0   & 0  & 0  &  0  & 1/2 &  0  & 0  & 0  \\
\hline
\end{tabular}
}
\begin{tabular}{|l|rrrr|}
\hline
$G_1 \ot G_1$ & $A_1$ & $T_1^1$ & $T_1^2$ & $T_1^3$ \\
\hline
$G_1^1 G_1^1$ &   0 & 1 & 0 & 0 \\
$G_1^1 G_1^2$ &  1/2& 0 &1/2& 0 \\
$G_1^2 G_1^1$ & -1/2& 0 &1/2& 0 \\
$G_1^2 G_1^2$ &   0 & 0 & 0 & 1 \\
\hline
\end{tabular}
\begin{tabular}{|l|rrrr|}
\hline
$G_1 \ot G_2$ & $A_2$ & $T_2^1$ & $T_2^2$ & $T_2^3$ \\
\hline
$G_1^1 G_2^1$ &  0 & 0 &  0 & -1 \\
$G_1^1 G_2^2$ & 1/2& 0 &-1/2&  0 \\
$G_1^2 G_2^1$ &-1/2& 0 &-1/2&  0 \\
$G_1^2 G_2^2$ &  0 & 1 &  0 &  0 \\
\hline
\end{tabular}
\begin{tabular}{|l|rrrr|}
\hline
$G_2 \ot G_2$ & $A_1$ & $T_1^1$ & $T_1^2$ & $T_1^3$ \\
\hline
$G_2^1 G_2^1$ &  0 & 1 & 0 &  0 \\
$G_2^1 G_2^2$ & 1/2& 0 &1/2&  0 \\
$G_2^2 G_2^1$ &-1/2& 0 &1/2&  0 \\
$G_2^2 G_2^2$ &  0 & 0 & 0 &  1 \\
\hline
\end{tabular}

\begin{tabular}{|l|rrrrrrrr|}
\hline $G_1 \ot H$ & $E^1$ & $E^2$ & $T_1^1$ & $T_1^2$ & $T_1^3$
            & $T_2^1$ & $T_2^2$ & $T_2^3$ \\
\hline
$G_1^1 H^1$ &  0 &1/2 &  0 &  0 &  0 &  0 & 1/2&  0 \\
$G_1^1 H^2$ &  0 &  0 &1/4 &  0 &  0 &3/4 &  0 &  0 \\
$G_1^1 H^3$ &1/2 &  0 &  0 &1/2 &  0 &  0 &  0 &  0 \\
$G_1^1 H^4$ &  0 &  0 &  0 &  0 &3/4 &  0 &  0 &1/4 \\
$G_1^2 H^1$ &  0 &  0 &-3/4&  0 &  0 &1/4 &  0 &  0 \\
$G_1^2 H^2$ &1/2 &  0 &  0 &-1/2&  0 &  0 &  0 &  0 \\
$G_1^2 H^3$ &  0 &  0 &  0 &  0 &-1/4&  0 &  0 &3/4 \\
$G_1^2 H^4$ &  0 &1/2 &  0 &  0 &  0 &  0 &-1/2&  0 \\
\hline
\end{tabular}
\begin{tabular}{|l|rrrrrrrr|}
\hline $G_2 \ot H$ & $E^1$ & $E^2$ & $T_1^1$ & $T_1^2$ & $T_1^3$
            & $T_2^1$ & $T_2^2$ & $T_2^3$ \\
\hline
$G_2^1 H^1$ &1/2 &  0 &  0 &1/2 &  0 &  0 &  0 &  0 \\
$G_2^1 H^2$ &  0 &  0 &  0 &  0 &-3/4&  0 &  0 &-1/4\\
$G_2^1 H^3$ &  0 &-1/2&  0 &  0 &  0 &  0 &-1/2&  0 \\
$G_2^1 H^4$ &  0 &  0 &1/4 &  0 &  0 &3/4 &  0 &  0 \\
$G_2^2 H^1$ &  0 &  0 &  0&   0 &-1/4&  0 &  0 &3/4 \\
$G_2^2 H^2$ &  0 &-1/2&  0 &  0 &  0 &  0 & 1/2&  0 \\
$G_2^2 H^3$ &  0 &  0 &3/4 &  0 &  0 &-1/4&  0 &  0 \\
$G_2^2 H^4$ &1/2 &  0 &  0 &-1/2&  0 &  0 &  0 &  0 \\
\hline
\end{tabular}
\vspace{-1mm} {\tabcolsep=0.3mm
\begin{tabular}{|l|rrrrrrrrrrrrrrrr|}
\hline $H \ot H$
   & $A_1$ & $A_2$ & $E^1$ & $E^2$ & $T_1^1$ & $T_1^2$ & $T_1^3$
   & $T_1^1$ & $T_1^2$ & $T_1^3$
   & $T_2^1$ & $T_2^2$ & $T_2^3$
   & $T_2^1$ & $T_2^2$ & $T_2^3$ \\
\hline
$H^1 H^1$ &  0&  0&  0&  0&  0&  0&  0&  0&  0& 25&  0&  0&  3&  0&  0&  0 \\
$H^1 H^2$ &  0&  1&  0&  1&  0&  0&  0&  0&  0&  0&  0& -1&  0&  0&  1&  0 \\
$H^1 H^3$ &  0&  0&  0&  0&  3&  0&  0&  3&  0&  0&  1&  0&  0&  1&  0&  0 \\
$H^1 H^4$ &  1&  0&  1&  0&  0&  9&  0&  0& -1&  0&  0&  0&  0&  0&  0&  0 \\
$H^2 H^1$ &  0&  1&  0& -1&  0&  0&  0&  0&  0&  0&  0& -1&  0&  0& -1&  0 \\
$H^2 H^2$ &  0&  0&  0&  0& -4&  0&  0&  9&  0&  0&  3&  0&  0&  0&  0&  0 \\
$H^2 H^3$ & -1&  0&  1&  0&  0& -1&  0&  0& -9&  0&  0&  0&  0&  0&  0&  0 \\
$H^2 H^4$ &  0&  0&  0&  0&  0&  0&  3&  0&  0&  3&  0&  0& -1&  0&  0&  1 \\
$H^3 H^1$ &  0&  0&  0&  0&  3&  0&  0&  3&  0&  0&  1&  0&  0& -1&  0&  0 \\
$H^3 H^2$ &  1&  0& -1&  0&  0& -1&  0&  0& -9&  0&  0&  0&  0&  0&  0&  0 \\
$H^3 H^3$ &  0&  0&  0&  0&  0&  0& -4&  0&  0&  9&  0&  0& -3&  0&  0&  0 \\
$H^3 H^4$ &  0& -1&  0&  1&  0&  0&  0&  0&  0&  0&  0& -1&  0&  0& -1&  0 \\
$H^4 H^1$ & -1&  0& -1&  0&  0&  9&  0&  0& -1&  0&  0&  0&  0&  0&  0&  0 \\
$H^4 H^2$ &  0&  0&  0&  0&  0&  0&  3&  0&  0&  3&  0&  0& -1&  0&  0& -1 \\
$H^4 H^3$ &  0& -1&  0& -1&  0&  0&  0&  0&  0&  0&  0& -1&  0&  0&  1&  0 \\
$H^4 H^4$ &  0&  0&  0&  0&  0&  0&  0& 25&  0&  0& -3&  0&  0&  0&  0&  0 \\
\hline {\it \small denom.\!}&
             4&  4&  4&  4& 10& 20& 10& 40& 20& 40&  8&  4&  8&  2&  4&  2 \\
\hline
\multicolumn{17}{l}{\it\small Numerators given in a column of this table should be divided }\\
\multicolumn{17}{l}{\it\small the listed denominator before taking the square root.}\\
\end{tabular}
}

\begin{thebibliography}{2}

\bibitem{Aoki03}
  S.~Aoki {\it et al.},
  Phys.\ Rev.\ D {\bf 67}, 034503 (2003).

 \bibitem{Weingarten93}
 F.~Butler, H.~Chen, J.~Sexton, A.~Vaccarino and D.~Weingarten,
 Phys.Rev.Lett. {\bf 70}, 2849 (1993).

\bibitem{Basak04}
  S.~Basak {\it et al.},
  Nucl.\ Phys.\ Proc.\ Suppl. {\bf 140}, 278 (2005).


\bibitem{Sato04}
  S.~Basak {\it et al.},
  Nucl.\ Phys.\ Proc.\ Suppl. {\bf 140}, 281 (2005).

\bibitem{BGR05}
  T. Burch {\it et al.}, Nucl. Phys. A{\bf 755}, 481 (2005).

\bibitem{BGR04}
 T. Burch {\it et al.}, Phys. Rev. D{\bf 70}, 054502 (2004).

\bibitem{Gockeler02}
  M.~G\"ockeler {\it et al.},
  Phys.\ Lett.\ B{\bf 532}, 63 (2002).

\bibitem{Kyoto03}
  Y.~Nemoto, N.~Nakajima, H.~Matsufuru, and H.~Suganuma,
  Phys.\ Rev.\ D {\bf 68}, 094505 (2003).

\bibitem{Sasaki05}
  K.~Sasaki and S.~Sasaki, hep-lat/0503026.

\bibitem{Sasaki02}
  S.~Sasaki, T.~Blum, and S.~Ohta,
  Phys.\ Rev.\ D {\bf 65}, 074503 (2003).

\bibitem{Adelaide03}
  W.~Melnitchouk {\it et al.},
  Phys.\ Rev.\ D {\bf 67}, 114506 (2003)

\bibitem{Regensburg03}
  D.~Br\"ommel {\it et al.},
  Phys.\ Rev.\ D {\bf 69}, 094513 (2004).

\bibitem{Leinweber03}
  J.~M.~Zanotti {\it et al.},
  Phys.\ Rev.\ D {\bf 68}, 054506 (2003).

\bibitem{michael85}
  C.~Michael, Nucl.\ Phys.\ B{\bf 259}, 58 (1985).

\bibitem{lw90}
  M.~L\"{u}scher and U.~Wolff, Nucl.\ Phys.\ B{\bf 339}, 222 (1990).

\bibitem{Morningstar99}
  C.~J.~Morningstar and M.~Peardon,
  Phys.\ Rev.\ D {\bf 60}, 034509 (1999).

\bibitem{Morningstar04}
  S.~Basak {\it et al.},
  Nucl.\ Phys.\ Proc.\ Suppl. {\bf 140}, 287 (2005).

\bibitem{Basak05}
  S.~Basak {\it et al.},
  submitted for publication.

\bibitem{Johnson}
  R.~C.~Johnson, Phys.\ Lett.\ B{\bf 114}, 147 (1982).

\bibitem{Elliott}
  J.~P.~Elliott and P.~G.~Dawber, {\it Symmetry in Physics}
  (Oxford University Press, New York 1979).

\bibitem{Altmann}
  S.~L.~Altmann and P.~Herzig, {\it Point-Group Theory Tables}
  (Oxford University Press, New York 1994).

\bibitem{Butler}
  P.~H.~Butler, {\it Point Group Symmetry Applications}
  (Prenum Press, New York 1981).

\bibitem{Mandula83}
  J.~Mandula, G.~Zweig, and J.~Govaerts,
  Nucl.\ Phys.\ B{\bf 228}, 91 (1983).

\bibitem{Mandula82}
  J.~Mandula and E.~Shpiz,
  Nucl.\ Phys.\ B{\bf 232}, 180 (1984).

\bibitem{Cracknell}
  S.~L.~Altmann and A.~P.~Cracknell,
  Reviews of Mordern Physics {\bf 37}, 1 (1965).

\bibitem{Dirl}
  R.~Dirl {\it et al.},
  Phys.\ Rev.\ B {\bf 32}, 788 (1985).

\bibitem{Wingate95}
  M.~Wingate, T.~DeGrand, S.~Collins, and U.~M.~Heller,
  Phys.\ Rev.\ D {\bf 52}, 307 (1995).

\bibitem{Lacock96}
  P.~Lacock, C.~Michael, P.~Boyle, and P.~Rowland,
  Phys. Rev. D {\bf 54}, 6997 (1996).



\bibitem{Alford96}
   M.~Alford, T.~Klassen and P.~Lepage, Nucl.\ Phys.\ Proc.\
   Suppl.\ {\bf 47}, 370 (1996).

\bibitem{ukqcd93}
  C.~R.~Allton {\it et al.},
  Phys.\ Rev. D {\bf 47}, 5128 (1993).

\bibitem{Guesken90}
  S.~Guesken, Nucl.\ Phys.\ Proc.\ Suppl. {\bf 17}, 361 (1990).


\bibitem{Albanese87}
 M.~Albanese {\it et al.},
  Phys.\ Lett.\ B{\bf 192}, 163 (1987).

\bibitem{Hasenfratz01}
  A.~Hasenfratz and F.~Knechtli,
  Phys.\ Rev.\ D {\bf 64}, 034504 (2001).

\bibitem{Peardon04}
  C.~Morningstar and M.~Peardon,
  Phys.\ Rev.\ D {\bf 69}, 054501 (2004).

\bibitem{Gammel71}
  J.~L.~Gammel, M.~T.~Menzel, and W.~R.~Wortman,
  Phys.\ Rev.\ D {\bf 3}, 2175 (1971).

\bibitem{Kubis72}
  J.~J.~Kubis,
  Phys.\ Rev.\ D {\bf 6}, 547 (1972).
\bibitem{PDG}
  S.~Eidelman \textit{et al}.,
  Phys.\ Lett.\ B{\bf 592}, 1 (2004).

\bibitem{IkuroThesis}
 I.~Sato, {\it Lattice QCD Simulations of Baryon Spectra and
 Development of Improved Interpolating Field Operators}, Ph.D.
 thesis submitted to the University of Maryland, August 2005
 (unpublished).

%
%
%
\end{thebibliography}
\end{document}